\documentclass[11pt]{article}
\usepackage{rotating}
\usepackage{amsfonts}
\usepackage{amsmath}
\usepackage{amsthm}
\usepackage{graphicx}
\usepackage{array, color}
\usepackage{bm}

\theoremstyle{definition}

\theoremstyle{remark}

\makeatletter\@addtoreset{equation}{section}
\renewcommand\theequation
{\ifnum \c@section>\z@ \thesection.\fi\@arabic\c@equation}
\makeatother

\begin{document}

\noindent{\Large \bf \textsf{Nearly Semiparametric Efficient Estimation of Quantile Regression }}

\vspace*{0.2in}

\noindent {\large \textsf{Kani C{\normalsize{\textsf{HEN}}},
 Yuanyuan L{\normalsize{\textsf{IN}}}, Zhanfeng W{\normalsize{\textsf{ANG}} and Zhiliang Y{\normalsize{\textsf{ING}} } }}

\vspace*{0.2in}

\noindent {\bf ABSTRACT}: As a competitive alternative to least squares regression, quantile regression is popular in analyzing
heterogenous data. For quantile regression model specified for one single quantile level $\tau$, major difficulties of semiparametric
efficient estimation are the unavailability of a parametric efficient score and the conditional density estimation.
In this paper, with the help of
the least favorable submodel technique,  we first derive the semiparametric efficient scores for linear quantile regression models that are assumed for
a single quantile level, multiple quantile levels and all the quantile levels in $(0,1)$ respectively.
Our main discovery is a one-step (nearly) semiparametric efficient estimation
for the regression coefficients of the quantile regression models assumed for multiple
quantile levels, which
has several advantages: it could be regarded as an optimal way to pool information across multiple/other
quantiles for efficiency gain;
it is computationally feasible and easy to implement, as the initial estimator is easily available; due to the nature
of quantile regression models under investigation,
the conditional density estimation is straightforward by plugging in an initial estimator.
The resulting estimator is proved to achieve the corresponding semiparametric efficiency lower bound under regularity conditions.
Numerical studies including simulations and an example of birth weight of children confirms that
the proposed estimator leads to higher efficiency compared with the Koenker-Bassett  quantile regression estimator
for all quantiles of interest.

%

%

%

%
%


 \begin{center}
{\textsf{1. \hspace{0.1in} INTRODUCTION}} \end{center}

\vspace{0.1in}

Quantile regression is a statistical methodology for the modeling
and inference of conditional quantile functions. Following Koenker and
Bassett (1978), we model the $\tau$th conditional quantile function of
$Y\in R$ given $X\in R^{p}$ as
\begin{eqnarray}
\label{m1}
Q_{Y|X}(\tau)= X^\top {\bm \beta}_\tau,
\end{eqnarray}
for certain specific $\tau\in (0,1)$, and ${\bm\beta}_\tau$ is $p$-vector usually
including an intercept.
Let $(x_i,y_i),~i=1,2,...,n$, be independent and identically distributed copies of $(X,Y)$.
For the $\tau$th quantile,
 the  classical Koenker-Bassett estimate of $\beta_\tau$, denoted as
 $\hat {\bm\beta}_\tau^c$, is obtained by minimizing the following
 objective function
\begin{eqnarray}
\label{obj1}
\sum_{i=1}^n \rho_\tau(y_i-x_i^\top {\bm \beta}_\tau),
\end{eqnarray}
over ${\bm \beta}_\tau$, where
$\rho_\tau(u)=u(\tau- I(u<0))$. The computation of $\hat{\bm \beta}^c_\tau$ is straightforward
with the help of linear programming.
There is vast literature on the estimation and inference for one or several percentile levels for model (\ref{m1}); see
 Yu and Jones (1998), He (1997),
  Koenker and Geling (2001), Koenker and Xiao (2002),  He and Zhu (2003), Koenker (2005),  Peng and Huang (2008),  Peng and Fine (2009),
 Bondell, Reich and Wang (2010),  Wang, Wu and Li (2012),  Jiang, Wang and Bondell (2013), He, Wang and Hong (2013), Kato (2011, 2012),
Zheng, Peng and He (2015), among many others.
 When there are commonality of quantile coefficients across multiple quantiles, the composite quantile regression (CQR) is proposed
to combine information shared across a number of quantiles to improve estimation efficiency; see Zou and Yuan (2008), Wang and Wang (2009),
Kai {\it et al.} (2001), Wang, Li and He (2012), Wang and Li (2013). But the novelty of CQR lies in the key assumption that there exist common covariate
effects across multiple quantile levels.
Recently,
important findings in Bayesian inference for quantile regression were reported in Yang and He (2011), Kim and Yang (2011) and Feng, Chen and He (2015).

Typically, model (\ref{m1}) can be expressed as the following linear
regression model
\begin{eqnarray}
\label{m1-2} Y=X^\top {\bm \beta}_\tau+\epsilon_\tau,
\end{eqnarray}
where the $\tau$th percentile of $\epsilon_\tau$ is assumed to be 0.
For specific $\tau$, under the independence assumption of $X$ and $\epsilon_\tau$,
it can be shown that $\hat {\bm \beta}_\tau^c$ is semiparametric efficient
by a straightforward argument to be discussed in section 2. As a special case, when $\tau=0.5$, the least absolute
deviation (LAD) is semiparametric efficient for model (\ref{m1-2}) with
the independence assumption of $X$ and $\epsilon_\tau$ (Zhou and Portnoy, 1998).
However, we point out that, without assuming independence of $X$ and $\epsilon_\tau$, $\hat{\bm \beta}_\tau^c$ is not semiparametric efficient and
the semiparametric
efficient estimation of model (\ref{m1}) or model(\ref{m1-2}) is indeed a sophisticated issue. The most difficult part is the estimation of the density of $\epsilon_\tau$ given $X$  in the semiparametric score function (Kato, 2014), which suffers from the curse of dimensionality.

When model (\ref{m1}) is specified for each $\tau \in (0,1)$,  following Portnoy (2003),
 we consider the quantile regression model
 \begin{eqnarray}
\label{qrm-2} Q_{Y|X}(\tau)=X^\top {\bm \beta}(\tau), \hskip 1cm \mbox{for all}\  \tau\in(0,1),
 \end{eqnarray}
where $Y$ and $X$ are the same as in model (\ref{m1}), and the regression parameter ${\bm \beta}(\tau)=(\beta_1(\tau),\beta_2(\tau),\cdots,\beta_p(\tau))^T$
is a function of $\tau$.  With the linearity assumption for all quantiles,
the true unknown function ${\bm \beta}_0(\tau)$ is suffice to describe the entire conditional distribution of $Y$ given $X$.
Important results on the estimation of the quantile process with survival data can be found in Portnoy (2003), Peng and Huang (2008).
Recently, there are some breakthroughs on Bayesian nonparametric regression
models on all quantiles; see
M$\ddot{u}$ller \& Quintana (2004), Dunson \& Taylor (2005) and Chung \& Dunson (2009),
 Reich {et al.}(2011), Qu \& Yoon (2015), etc. To summarize,
there are two main approaches for the estimation of quantile process: linear interpolation
 and basis representation. The linear interpolation approach consists of two steps: the first step is to estimate the quantile regression coefficients separately
at certain proper grid of $\tau$-values, and the second step is to interpolate linearly between grid values or apply rearrangement.
For the basis representation method, the quantile function is
represented by basis functions or some specific functions after transformation. Nevertheless, both methods reviewed above are in Bayesian
framework and their theoretical properties remain unclear.

To the best of our knowledge, there is no specific construction of a semiparametric efficient estimate of
 ${\bm \beta}(\tau)$ of model (\ref{qrm-2}) in the literature.  We point out that for model (\ref{qrm-2}), the likelihood function is
$\prod_{i=1}^n 1/\{x_i^\top \bm {\dot \beta}(\tau_i) \}$ where $y_i=x_i^\top \bm \beta(\tau_i)$
and $\dot {\bm \beta}(\cdot)$ is the derivative of ${\bm \beta}(\cdot)$. However,
the maximum likelihood method as in Zeng and Lin (2006,2007) involves enormous technical/numerical difficulty.
In our view, one of the main reasons lies
in the nature of model (\ref{qrm-2}) that the quantile process $\bm \beta(\cdot)$ and the nuisance parameter $f_{Y|X}$ are not separable.
The numerical maximization of the estimated likelihood subject to $n$ constrains $y_i=x_i^\top {\bm \beta}(\tau_i)$ is rather unstable and wild. The numerical difficulties
here are in the same spirit as that in numerically searching for the maximum likelihood estimation (MLE) of $\theta$ for Uniform$[0,\theta]$, where the solutions would often go to the boundary. Moreover, due to data sparsity, the estimated $\dot {\bm \beta}(\tau)$ or $\dot {\bm \beta}(\tau)$ would be unstable when $\tau$ is close to 0 or 1.

In view of the technical/numerical complications involved in the semiparametric efficient estimation
 of ${\bm \beta}(\tau)$ in model (\ref{qrm-2}), 
we thus take one step back and consider the following quantile regression model
\begin{eqnarray}\label{qrm}
Q_{Y|X}(\tau_l)=X^\top {\bm \beta}(\tau_l), \hskip 1cm \mbox{for all}\ l=1, 2, \ldots, L,
\end{eqnarray}
where $0<\tau_1<\tau_2<\cdots<\tau_L<1$. Model (\ref{qrm}) is intermediate of model (\ref{m1}) and
model (\ref{qrm-2}). With the explicit expression of the semiparametric efficient score function
of ${\bm \beta}(\tau_k)$, $k=1,2,\ldots, L$,
derived by the least favorable submodel technique in section 2, we propose a one-step estimation with
the estimated  score function, that leads to the semiparametric efficient estimation
of ${\bm \beta}(\tau_k)$.  The proposed procedure is numerically doable and stable.
Most importantly, one can show that when the maximum space of $\{\tau_l-\tau_{l-1},l=1,2,\ldots,L+1\}$ tends to 0,
the semiparametric efficient score of model (\ref{qrm}) approaches to that of
model (\ref{qrm-2}). As the impetus for this work was to pursue semiparametric efficient estimation of
 ${\bm \beta}(\tau)$ in model (\ref{qrm-2}), theoretically, one can use efficient estimator of
${\bm \beta}(\tau_k)$ with model (\ref{qrm}) to approximate that of model (\ref{qrm-2}). Hence, we refer the proposed procedure
as {\it nearly semiparametric efficient estimation for quantile regression}.

The rest of the paper is organized as follows. Section 2 introduces the model and
the proposed estimation with detailed discussions.
Extensive simulation studies with supportive evidence are demonstrated in section 3.
In section 4, the proposed method is illustrated using a real data of birth weight of children from the National Center for Health Statistics.
All technical derivation and proofs are in Appendix.

%
%

\vspace{3ex}

\begin{center}
\textsf{2. \hspace{0.1in}  METHODOLOGIES AND MAIN RESULTS}
\end{center}
\vspace{0.1in}

First,  consider model (\ref{qrm}), by the definition of quantile,
\begin{eqnarray}
F_{Y|X}(Q_{Y|X}(\tau_l))=\tau_l \ \ \Rightarrow \ \ F_{Y|X}(X^\top{\bm \beta}(\tau_l))=\tau_l,~~
~~l=1,2,\cdots,L,\label{cdf}
\end{eqnarray}
where $F_{Y|X}$ is the cumulative distribution function of $Y$ given $X$. Let $f_{Y|X}(t)$ be the density function of $Y$ conditional on $X$.
Let ${\bm \beta}_{0}(\tau_l)=(\beta_{10}(\tau_l),\cdots,\beta_{p0}(\tau_l))^\top$ be the true value of
${\bm \beta}(\tau_l)=(\beta_{1}(\tau_l),\cdots,\beta_{p}(\tau_l))^\top$. By the nature of quantile regression model,
 $x^\top{\bm \beta}(\tau_l)$ is $\tau_l$-quantile of $Y$ given $X=x$.
Without loss of generality, we assume that $x^\top{\bm \beta}(\tau_1)<x^\top{\bm \beta}(\tau_2)<\cdots<x^\top{\bm \beta}(\tau_L)$.

\vspace{0.1in}

\noindent {\it 2.1.  Semiparametric efficient scores. }

In quantile regression, estimation of the quantile regression coefficient or the quantile process is inseparably linked
to the nuisance parameter, the conditional density function. In such a case,  the least favorable submodel
method (Kato, 2014) plays a role to derive a semiparametric efficient score function of ${\bm \beta}(\tau_l),~l=1,...,L$ of model (\ref{qrm})  and
their variance lower bound. It is known that the least favorable submodel technique is to reduce a high dimensional
problem to a problem involving a finite-dimensional `` least favorable submodel"; see Begun {\it et al.}(1983), Bickel {\it et al.}(1993), among others.
 Following section 25.4 in van der Vaart (1998),
we begin with the construction of a parametric submodel of model (\ref{qrm}) based on the cumulative distribution function
with parameter $\theta$ in a neighborhood of 0,
\begin{eqnarray}
\tilde F_{Y|X}(t;\theta)=F_{Y|X}(t)+\theta G_{Y|X}(t),
\label{psub}
\end{eqnarray}
where $G_{Y|X}(t)$ is a function of $t$ satisfying certain conditions.
Differentiating (\ref{psub}) we get
\begin{eqnarray}
\tilde f_{Y|X}(t;\theta)=f_{Y|X}(t)+\theta g_{Y|X}(t),
\label{sub-model1}
\end{eqnarray}
where $\tilde f_{Y|X}(t;\theta)$, $f_{Y|X}(t)$ and $g_{Y|X}(t)$ are derivatives of $\tilde F_{Y|X}(t;\theta)$, $F_{Y|X}(t)$
and $G_{Y|X}(t)$ respectively.
To guarantee  $\tilde f_{Y|X}(t;\theta)$ is a density function
for all $\theta$, the first restriction of  $G_{Y|X}(t)$ is that $$\int_{-\infty}^{+\infty} g_{Y|X}(u) du=0.$$
Moreover, under model (\ref{qrm}),
let $X^\top{\bm{\beta}}(\tau_l;\theta)$ be the $\tau_l$ quantile of $\tilde F_{Y|X}(t;\theta)$ and
$X^\top{\bm{\beta}}(\tau_l;0)=X^\top{\bm{\beta}}_{0}(\tau_l)$,  for $l=1,2,\ldots,L$.
Hence, we have the identity
$\tau_l=\tilde{F}_{Y|X}(x^\top{\bm{\beta}}(\tau_l;\theta);\theta)$.
By a Taylor expansion of the right hand side of this identity as a function of $\theta$
in a neighborhood of 0,  we
obtain the second restriction that $$G_{Y|X}(X^\top{\bm \beta}_{0}(\tau_l))=-f_{Y|X}(X^\top{\bm \beta}_{0}(\tau_l))X^\top{\bm d}(\tau_l),$$ for $l=1,2
\ldots,L$, where ${\bm d}(\tau_l)$ is the derivative of ${\bm{\beta}}(\tau_l;\theta)$ at $\theta=0$.
Clearly,
the derivative of log-likelihood of $\theta$ based on the density function $\tilde f_{Y|X}(t)$ at $\theta=0$
is  ${g_{Y|X}(t)}/{f_{Y|X}(t)} $, denoted as $\xi$.
By the information theory in  Bickel {et al.}(1993),
we are able to approximate the least favorable submodel by searching for the lower bound of
$E(\xi^2)$, which as a result would lead to the semiparametric efficient score. We defer the details to Appendix I.
The resulting semiparametric efficient score of ${\bm \beta}(\tau_k)$ can be regarded as an optimal way to combine information from
all the quantile levels $\tau_1,...,\tau_L$.

Let ${\bm U}={\bm B} {\bm A} {\bm B}^\top$ and ${\bm W}$ be a $pL\times pL$ diagonal matrix
with diagonal elements being the reciprocal of diagonal of matrix ${\bm U}^{-1}$,
where ${\bm A}$ and ${\bm B}$ are defined in (\ref{Adef}) and (\ref{Bdef}) in Appendix I. Set $({\bm u}_1,{\bm u}_2,\cdots,{\bm u}_{pL})={\bm U}^{-1}{\bm W}$, where ${\bm u}_i$ is a vector with length $pL$.
The following proposition presents the semiparametric efficient score of ${\bm \beta}(\tau_k)$, $1\leq k\leq L$ and their variance
lower bound.

\vspace{0.1in}

\noindent \textbf{Proposition 1.} \emph{For model (\ref{qrm}), the semiparametric efficient score of ${\bm \beta}(\tau_k)$,
$1\leq k\leq L$, is
\begin{eqnarray}\label{eq1-1}
S_{k}(y,x)=&\sum_{l=1}^{L+1}\frac{f_{Y|X}(x^\top{\bm \beta}(\tau_{l-1}))
x^\top{\bm D}_{l-1}-f_{Y|X}(x^\top{\bm \beta}(\tau_{l}))
x^\top{\bm D}_{l}}{\tau_l-\tau_{l-1}}\Big[I\{x^\top{\bm \beta}(\tau_{l-1})<y<x^\top{\bm \beta}(\tau_{l})\}\nonumber\\
&-(\tau_l-\tau_{l-1})\Big].
\end{eqnarray}
Moreover, for the estimate of the $j-$th component of ${\bm \beta}(\tau_k)$, its variance has a lower bound
\begin{eqnarray}\label{eq1-2}
\sigma_{kj}^2=\frac{1}{{\bm u}_{kj}^\top {\bm U} {\bm u}_{kj}},~~~~ j=1,2,\ldots,p,\nonumber
\end{eqnarray}
where ${\bm D}_l$ is $p\times p$ matrix, $l=0,1,\ldots,L+1$, ${\bm D}_0={\bm D}_{L+1}=0$, $({\bm D}_1,{\bm D}_2,\ldots,{\bm D}_L)=({\bm u}_{k1},{\bm u}_{k2},\ldots,{\bm u}_{kp})^\top$;
 $f_{Y|X}(x^\top{\bm \beta}_{0}(0))=f_{Y|X}(x^\top{\bm \beta}_{0}(1))=0$; 
 ${\bm \beta}_{0}(0)=-\infty$, ${\bm \beta}_{0}(1)=+\infty$; $\tau_0=0$ and $\tau_{L+1}=1$.
}

\vspace{0.1in}

{\it Remark 1.} When $L=1$, model (\ref{qrm}) reduces to model (\ref{m1}) for one single quantile point $\tau$.
By Proposition 1, the semiparametric efficient score of ${\bm \beta}_{\tau}$ in model (\ref{m1}) is
\begin{eqnarray}\label{sco-L1}
S(y,x)=f_{Y|X}(x^\top{\bm \beta}_{\tau})\frac{1}{(1-\tau)\tau}\left\{\tau-
I(y<x^\top{\bm \beta}_{\tau})\right\}{\bm D}^\top x.
\end{eqnarray}
By the definitions of ${\bm U}$ and ${\bm W}$, ${\bm D}={\bm U}^{-1}{\bm W}$ is a constant matrix not depending on random variable $X$. For the corresponding linear model (\ref{m1-2}),  under the assumption that  the $\tau$-quantile of $\epsilon_{\tau}$ is $0$ and the error term
$\epsilon_{\tau}$ is independent of covariate $X$,  $f_{Y|X}(x^\top{\bm \beta}_{\tau})=f_{Y-X^\top{\bm \beta}_{\tau}|X}(0)=f_{\epsilon_{\tau}|X}(0)$ is also not relevant to $X$. In this case, the efficient score in (\ref{sco-L1}) is exactly  the efficient score in classical quantile regression model specified at a single quantile level,  such as the least absolute deviation estimate (LAD) for median regression; see Zhou and Portnoy (1998) and  Kato (2014).  However,
without the crucial independence assumption of $X$ and $\epsilon_\tau$, as  conventional quantile regression models
allows heterogeneity, the distribution of $\epsilon_{\tau}$ depends on $X$ implying $f_{Y|X}(x^\top{\bm \beta}_{\tau})$ also depends on $x$. As a result,  the Koenker-Bassett estimate is not semiparametric efficient.

{\it Remark 2.} When $L\rightarrow \infty$ and the maximum space of $\{\tau_l-\tau_{l-1},l=1,2,\ldots,L+1\}$ tends to 0, model (\ref{qrm}) approaches model (\ref{qrm-2}).
Next, we intend to show that the semiparametric efficient score  (\ref{eq1-1}) of ${\bm \beta}(\tau_k)$ approaches that of model (\ref{qrm-2}) as $L\to \infty$. In fact, for
 the $j$-th component of ${\bm \beta}(\tau_k)$, a similar calculation as that of (\ref{eq1-1}) reveals that
 semiparametric efficient score of ${\beta}_j(\tau_k)$ in model (\ref{qrm-2})
is
\begin{align}\label{m-s2}
S^*_{kj}(y,x)=&-\frac{\partial (f_{Y|X}(x^\top{\bm \beta}(\tau))x^\top{\bm d}(\tau))}{\partial \tau}
\doteq -\dot{\Big\{
f_{Y|X}(x^\top{\bm \beta}(\tau))x^\top{\bm d}(\tau)\Big\}},
\end{align}
where ${\bm d}(\tau)=[d_1(\tau),...,d_p(\tau)]^\top$ is a minimizer of
\begin{align}\label{mqd-2}
E[\{S^*_{kj}(Y,X)\}^2]=\int_{-\infty}^{\infty}\left[\dot{\Big\{
f_{Y|X}(X^\top{\bm \beta}(\tau))X^\top{\bm d}(\tau)\Big\}}\bigg |_{t=X^\top{\bm \beta}(\tau)}\right]^2dt,
\end{align}
subject to ${d}_j(\tau_k)=1$. We defer the detailed derivations of this finding in Appendix II. We point out that, it is
infeasible to pursue the semiparametric efficient estimation of ${\beta}_j(\tau_k)$ in model (\ref{qrm-2})  based on
(\ref{m-s2}), as the numerical minimization of  (\ref{mqd-2}) is intractable.
Fortunately,  the semiparametric efficient score  of ${ \beta}_{j}(\tau_k)$ in (\ref{eq1-1}) can be rewritten as,
\begin{align}
S_{kj}(y,x)
=\sum_{l=1}^{L+1}\frac{f_{Y|X}(x^\top{\bm \beta}(\tau_{l-1}))
x^\top{\bm d}(\tau_{l-1})-f_{Y|X}(x^\top{\bm \beta}(\tau_{l}))
x^\top{\bm d}(\tau_{l})}{\tau_l-\tau_{l-1}}\nonumber\\
I\{x^\top{\bm \beta}(\tau_{l-1})<y<x^\top{\bm \beta}(\tau_{l})\},\label{m-s1}
\end{align}
where ${\bm d}=[{\bm d}(\tau_1)^\top,...,{\bm d}(\tau_L)^\top]^\top$ is a minimizer of the quadratic form
$E[\{S_{kj}(Y,X)\}^2]\equiv {\bm d}^\top {\bm B} {\bm A} {\bm B}^\top {\bm d}$  subject to $d_{j}(\tau_k)=1$.
It is straightforward to check that
\begin{align}
S_{kj}(y,x)\to S^*_{kj}(y,x)~~{\mbox{and}}~~E[\{S_{kj}(Y,X)\}^2] \to E[\{S^*_{kj}(Y,X)\}^2]
\end{align}
as $L\to \infty$. This finding motivates us to use the efficient score in (\ref{eq1-1}) to approximate the efficient score in (\ref{m-s2}), which leads to a nearly semiparametric efficient estimator of ${\bm \beta}(\tau_k)$ in model (\ref{qrm-2}).

{\it Remark 3. }  The key idea of this work is to borrow information across quantiles and search for the most efficient estimation. This remark provides more insights in this idea.
Intuitively, for certain quantile level $\tau_k$,  the estimation of ${\bm \beta}(\tau_k)$ in traditional quantile regression does not depend on the information
on $Y$ at other quantiles $\{\tau_i,~i\neq k\}$, especially those quantiles far away from $\tau_k$. The intuition is true when the number of covariates (including an intercept term) is 1, that is $p=1$.  For this special case, one can  rewrite (\ref{m-s1}) as
\begin{align}
\label{rm3}
S_{kj}(y,x)=&\left\{\frac{{{\bm d}(\tau_{k})}/{\dot{\bm \beta}(\tau_{k})}-{{\bm d}(\tau_{k+1})}/{\dot{\bm \beta}(\tau_{k+1})}}{\tau_{k+1}-\tau_{k}}-\frac{{{\bm d}(\tau_{k-1})}/{\dot{\bm \beta}(\tau_{k-1})}-{{\bm d}(\tau_{k})}/{\dot{\bm \beta}(\tau_{k})}}{\tau_k-\tau_{k-1}}\right\}\nonumber\\
&[\tau_k-I\{y<{\bm \beta}(\tau_{k})x\}],
\end{align}
 from which one can see that $S_{kj}(y,x)$ is not relevant to the model information at other quantiles $\{\tau_l, l\neq k\}$. Appendix III contains the proofs of $(\ref{rm3})$.
 In other words, for model (\ref{qrm}) with $p=1$, the semiparametric efficiency for the estimation of $\beta_j(\tau_k)$ can be achieved using only the information at $\tau_k$. However, besides an intercept, there is generally at least one covariate in the model,  namely $p\geq 2$. Hence,  the efficient estimator of $\beta_j(\tau_k)$ generally depends on the information at other quantiles. In view of this fact,
borrowing information across other quantiles via the efficient score $(\ref{rm3})$ is able to improve the estimation efficiency of ${\bm \beta}(\tau_k)$ when $p\geq 2$. In addition, Proposition 1 tells that the variance of estimates of $\beta_j(\tau_k)$ have a  lower bound $\sigma_{kj}^2$.

For illustration, we consider a toy example for model (\ref{qrm}) with $L=2$.
To estimate $\beta_1(\tau_1)$,  if we use only the model information at single quantile $\tau_1$ and ignore the information at $\tau_2$,
then
\begin{eqnarray}
E[\{S_{11}(Y,X)\}^2]=E\left[\frac{1}{\tau_1(1-\tau_1)}\left\{f_{Y|X}(X^\top{\bm\beta}(\tau_1))
\right\}^2\left\{X^\top{\bm d}(\tau_1)\right\}^2\right]\doteq E(Q_1).\nonumber
\end{eqnarray}
 On the other hand,
by incorporating the model information at $\tau_2$ for the estimation of $\beta_1(\tau_1)$, we have shown in Appendix IV that
\begin{align}
&E[\{S_{11}(Y,X)\}^2]\nonumber \\
 =&E\Big[\frac{1}{\tau_1}\left\{f_{Y|X}(X^\top{\bm\beta}(\tau_1))\right\}^2
\left\{X^\top{\bm d}(\tau_1)\right\}^2+\frac{1}{1-\tau_2}\left\{f_{Y|X}(X^\top{\bm\beta}(\tau_2))\right\}^2\left\{X^\top{\bm d}(\tau_2)\right\}^2\nonumber\\
&+
\frac{1}{\tau_2-\tau_1}\left\{f_{Y|X}(X^\top{\bm \beta}(\tau_1))X^\top{\bm d}(\tau_1)-
f_{Y|X}(X^\top{\bm\beta}(\tau_2))X^\top{\bm d}(\tau_2)\right\}^2\Big] \nonumber \\
\doteq & E(Q_2).
\end{align}
Most importantly, we have shown $Q_2-Q_1>0$ 
 which leads to $E(Q_2)-E(Q_1)>0$. In summary, our theoretical analysis validates that combining information across quantiles can generally
reduce the variance of the estimate of ${\bm \beta}(\tau_k)$.

\vspace{0.1in}

\noindent {\it 2.2. The nearly semiparametric efficient estimation.}

In this subsection, we introduce the proposed nearly semiparametric efficient estimation procedure for the regression coefficients
of mode (\ref{qrm-2}). As discussed earlier, we make use of the score (\ref{eq1-1}) in the construction of the proposed estimator.
Since (\ref{eq1-1}) involves the density function of $Y$ given $X$,
we need to find an appropriate estimate of $f_{Y|X}(x^\top{\bm\beta}(\tau_l))$, $l=1,2,\ldots,L$. Recall that
\begin{eqnarray}
f_{Y|X}(X^\top{\bm\beta}(\tau_l))=\frac{1}{X^\top\dot{\bm\beta}(\tau_l)}, ~~~~~l=1,2,\ldots,L. \label{densest}
\end{eqnarray}
Hence, instead of estimating the conditional density function directly, we estimate $\dot{\bm\beta}(\tau_l)$.
 A natural estimate of $\dot{\bm\beta}(\tau_l)$ is
$\hat{\dot{\bm\beta}}(\tau_l)=\{\hat{\bm \beta}^c(\tau_l+h)-\hat{\bm \beta}^c(\tau_l-h)\}/(2h)$,
where $\hat{\bm \beta}^c(\tau)$ is the Koenker-Bassett estimate of ${\bm \beta}(\tau)$ by minimizing (\ref{obj1}) and $h$ is the bandwidth.
Thus, the density function $f_{Y|X}(X^\top{\bm\beta}(\tau_l))$ can be estimated by
${1}/{X^\top {\hat{\dot{\bm\beta}}}(\tau_l)}$ for $l=1,2,\ldots,L$.
Next, we define the proposed one-step estimator of ${\bm \beta}(\tau_{k})$, denoted by $\hat{\bm \beta}(\tau_{k})$,  as
\begin{eqnarray}\label{onestep}
\hat{\beta}_{j}(\tau_k)=\hat{ \beta}^c_{j}(\tau_k)+\hat \sigma^2_{kj}\frac{\sum_{i=1}^n \hat S_{kj}(y_i,x_i)}{n},
~~j=1,2,\ldots,p,
\end{eqnarray}
where $\hat S_{kj}(y,x)$ is the $j$-th component of the estimated score $\hat {\bm S}_{k}(y,x)$ by plugging $\hat{\dot{\bm\beta}}(\tau_l)$ and $\hat{ \bm \beta}^c (\tau_l)$, $l=1,\ldots, L$, into (\ref{eq1-1}), $\hat \sigma_{kj}^2$ is the estimated variance lower bound by plugging $\hat{\dot{\bm\beta}}(\tau_l)$ and $\hat{ \bm \beta}^c (\tau_l)$, $l=1,\ldots, L$, into $\sigma_{kj}^2$ in Proposition 1.
Under regularity conditions given in Appendix V, the resulting estimate of $\hat{\beta}_{j}(\tau_k)$ can be proved to achieve the semiparametric efficiency lower bound. The following theorem presents the main results.

\vspace{0.1in}

\noindent \textbf{Theorem 1.} \emph{Assume model (\ref{qrm-2}) and conditions $(1)-(3)$ in Appendix V hold. Then,
for $j=1,2,\ldots,p$ and $k=1,2,\ldots,L$,
\begin{eqnarray}\label{Normality}
\sqrt{n}\left\{\hat{\beta}_{j}(\tau_k)-{\beta}_{0j}(\tau_k)\right\}\to N(0,~\sigma_{kj}^2)
\end{eqnarray}
in distribution as $n\to \infty$, where ${\beta}_{0j}(\tau_k)$ is the $j$-th component of ${\bm \beta}_{0}(\tau_k)$. Moreover, the
asymptotic variance of $\hat{\beta}_{j}(\tau_k)$ achieves the semiparametric efficiency bound $\sigma_{kj}^2$.
}

The implementation of the one-step estimation is as follows: for each $k=1,\cdots,L$, $j=1,\cdots,p$,\\
{\it Step 1. } For each $l=1,\cdots,L$,  compute the initial estimator $\hat{\bm \beta}^c(\tau_l)$;\\
{\it Step 2.}  For each $l=1,\cdots,L$,  calculate ${\hat{\dot{\bm\beta}}(\tau_l)}$ and  the conditional density function $f_{Y|X}({x_i^\top\bm \beta}(\tau_{l}))$ is ${1}/{x_i^\top \hat{\dot{\bm \beta}}(\tau_l)}$; \\
{\it Step 3. }  Compute $\hat S_{kj}(y,x)$ and $\hat \sigma_{kj}^2$ by plugging the initial estimator in step 1 and the estimated density in step 2 into $S_{kj}(y,x)$ and $\sigma_{kj}^2$; \\
{\it Step 4.}  Obtain $\hat{\beta}_j(\tau_{k})$ according to (\ref{onestep}).

{\it Remark 4. } Actually, in the above one-step estimation, we only need to estimate the conditional density function  $f_{Y|X}(x^\top{\bm\beta}(\tau_l))$ at quantile  levels $\{\tau_l,l=1,\ldots,L\}$.
 In this regard, we only need to assume the linear quantile regression model is specified in a neighborhood of each $\tau_l$, $l=1,\ldots,L$, and do not need
 to assume a linear quantile regression model for all $\tau\in(0,1)$.

\vspace{3ex}

\begin{center}
\textsf{3. \hspace{0.1in} SIMULATION STUDIES}
\end{center}
\vspace{0.1in}

Simulations are conducted to evaluate the performance of our proposed
method. In the simulation, for a quantile level $\tau_k$ of interest,
 we consider three methods for the estimation of $\hat{\beta}_{j}(\tau_k)$: the Koenker-Bassett quantile
estimate $\hat {\bm\beta}_\tau^c$, denoted by TQE; the proposed one-step estimate based on
the semiparametric efficient score of ${\bm \beta}(\tau_k)$, referred as EFF; the one-step estimate based on the score function (\ref{sco-L1})
ignoring the model information at other quantiles, referred as (SEF).
The simulated data is generated from the following quantile regression model with two
covariates,
\begin{eqnarray}\label{simu-qrm}
Q_{Y|X}(\tau)=X_1\beta_1(\tau)+X_2\beta_2(\tau),
\end{eqnarray}
where $\beta_1(\tau)$ and $\beta_2(\tau)$ takes each of the following 5 forms: \\
$M1:$ $\beta_1(\tau)=2$ and $\beta_2(\tau)=1+\Phi^{-1}(\tau)$;\\
$M2:$ $\beta_1(\tau)=2+\Phi^{-1}(\tau)$ and $\beta_2(\tau)=2+\Phi^{-1}(\tau)$;\\
$M3:$ $\beta_1(\tau)=2$ and $\beta_2(\tau)=1+\log\{\tau/(1-\tau)\}$;\\
$M4:$ $\beta_1(\tau)=2$ and $\beta_2(\tau)=1+\tan\{\pi*(\tau-0.5)\}$;\\
$M5:$ $\beta_1(\tau)=1+\log\{\tau/(1-\tau)\}$ and $\beta_2(\tau)=2+\tan\{\pi*(\tau-0.5)\}$.\\
The covariate $X_1$ is constant $1$  for $M1$, $M3$ and $M4$,  and it follows log-normal distribution for $M2$ and $M5$.
Another covariate $X_2$ follows log-normal distribution for all cases.
In particular, model (\ref{simu-qrm}) with
cases $M1$ and $M2$ are equivalent to
\begin{eqnarray}
Y=2+ X_2+X_2 \epsilon,\nonumber
\end{eqnarray}
and
\begin{eqnarray}
Y=2+2 X_2+(1+X_2)\epsilon,\nonumber
\end{eqnarray}
respectively, where $\epsilon$ follows the standard normal distribution.
The sample size $n=1000$ and 2000. All simulations are repeated 1000 times.

We first consider the two quantiles $0.5$ and $0.7$. The simulation results are summarized in
 Table 1. One can see that the parameter estimates
are generally unbiased. In all configurations, EFF has the smallest standard deviation (SD) compared with TQE and SEF.
And SEF have much smaller SD compared to TQE. For example, for case M3 and $n=1000$,  the ratio of the standard deviations of TQE and EFF ranges from  $1.343$ to $2.214$. And the ratio of the standard deviations of SEF and EFF ranges from $1.026$ to $1.062$. In other words,  EFF improves efficiency of TQE for at least $80\%$ and it improves efficiency of the SEF for around $5\%$ to 12\%, which confirms our theoretical findings.

\begin{table}[!h]
\tabcolsep=1.5pt\fontsize{8}{10}
\selectfont
\begin{center}
\caption {Simulation results for five models with quantiles 0.5 and 0.7.}\label{tab1}
\vskip 10pt
\begin{tabular}{cccc cccc}
\hline
&& &\multicolumn{2}{c}{$\tau=0.5$} & & \multicolumn{2}{c}{$\tau=0.7$}\\
\cline{4-5} \cline{7-8}
Model&$n$&&$\beta_1(\tau)$&$\beta_2(\tau)$& &$\beta_1(\tau)$&$\beta_2(\tau)$\\
\cline{2-8}

M1

&& True            & 2 & 1 && 2 & 1.5244 \\

&1000& TQE             &2.0007(0.0512)&0.9974(0.0899)&&2.0031(0.0547)&1.5195(0.0961)\\
&& SEF             &2.0009(0.0238)&0.9968(0.0547)&&2.0050(0.0265)&1.5149(0.0560)\\
&& EFF &2.0015(0.0227)&0.9959(0.0533)&&2.0009(0.0247)&1.5200(0.0529)\\

\cline{3-8}
&2000& TQE             &1.9992(0.0365)&1.0010(0.0652)&&2.0023(0.0370)&1.5213(0.0653)\\
&& SEF             &2.0002(0.0159)&0.9993(0.0361)&&2.0034(0.0174)&1.5190(0.0376)\\
&& EFF &2.0002(0.0145)&0.9992(0.0352)&&2.0006(0.0150)&1.5224(0.0365)\\

\cline{2-8}

M2

&& True            & 2 & 2 && 2.5244 & 2.5244 \\
&1000& TQE             &1.9976(0.1192)&1.9987(0.1155)&&2.5240(0.1244)&2.5206(0.1229)\\
&& SEF             &1.9989(0.0896)&1.9981(0.0875)&&2.5228(0.0891)&2.5209(0.0903)\\
&& EFF &1.9985(0.0881)&1.9982(0.0870)&&2.5239(0.0883)&2.5205(0.0881)\\
\cline{3-8}

&2000& TQE             &1.9980(0.0834)&2.0022(0.0844)&&2.5230(0.0877)&2.5225(0.0833)\\
&& SEF             &1.9990(0.0617)&2.0003(0.0614)&&2.5232(0.0631)&2.5236(0.0605)\\
&& EFF &1.9988(0.0608)&2.0002(0.0608)&&2.5240(0.0624)&2.5228(0.0602)\\
\cline{2-8}
M3

&& True            & 2 & 1 && 2 & 1.8473 \\

&1000& TQE             &2.0011(0.0822)&0.9958(0.1437)&&2.0055(0.0907)&1.8397(0.1592)\\
&& SEF             &2.0014(0.0381)&0.9949(0.0874)&&2.0094(0.0445)&1.8305(0.0929)\\
&& EFF &2.0021(0.0365)&0.9938(0.0852)&&2.0019(0.0420)&1.8400(0.0875)\\
\cline{3-8}

&2000& TQE             &1.9987(0.0585)&1.0017(0.1042)&&2.0040(0.0615)&1.8424(0.1082)\\
&& SEF             &2.0003(0.0256)&0.9990(0.0575)&&2.0061(0.0290)&1.8378(0.0622)\\
&& EFF &2.0002(0.0230)&0.9990(0.0561)&&2.0012(0.0250)&1.8436(0.0607)\\
\cline{2-8}

M4

&& True            & 2 & 1 && 2 & 1.7265 \\

&1000& TQE             &2.0009(0.0669)&0.9966(0.1144)&&2.0083(0.0930)&1.7221(0.1621)\\
&& SEF             &2.0014(0.0316)&0.9955(0.0699)&&2.0166(0.0491)&1.7015(0.0952)\\
&& EFF &2.0023(0.0287)&0.9945(0.0677)&&2.0041(0.0480)&1.7172(0.0925)\\
\cline{3-8}

&2000& TQE             &1.9990(0.0469)&1.0013(0.0824)&&2.0057(0.0629)&1.7228(0.1097)\\
&& SEF             &2.0002(0.0207)&0.9993(0.0461)&&2.0103(0.0327)&1.7118(0.0646)\\
&& EFF &2.0005(0.0188)&0.9988(0.0449)&&2.0016(0.0289)&1.7227(0.0628)\\
\cline{2-8}

M5&& True            & 1 & 2 && 1.8473 & 2.7265 \\

&1000& TQE             &0.9964(0.1797)&1.9982(0.1555)&&1.8467(0.2073)&2.7277(0.2072)\\
&& SEF             &0.9979(0.1344)&1.9972(0.1179)&&1.8440(0.1488)&2.7214(0.1510)\\
&& EFF &0.9971(0.1315)&1.9984(0.1173)&&1.8449(0.1465)&2.7250(0.1474)\\
\cline{3-8}
&2000& TQE             &0.9973(0.1258)&2.003(0.1139)&&1.8449(0.1459)&2.7268(0.1396)\\
&& SEF             &0.9987(0.0921)&2.0006(0.0831)&&1.8448(0.1052)&2.7260(0.1011)\\
&& EFF &0.9982(0.0911)&2.0004(0.0817)&&1.8462(0.1039)&2.7264(0.1004)\\

\hline
\multicolumn{8}{l}{$^*$ Standard deviations are in parentheses. }
\end{tabular}
\end{center}
\end{table}

In addition, we also compare the numerical performance of the three methods with quantiles $0.5$ and $0.9$, a higher quantile.
 Table 2 reports the estimation results for the 5 cases, from which similar conclusion to that of $\tau=0.5$ and $0.7$ can be drawn.
 Specially, EFF has the smallest standard erros and SEF is more efficient than TQE. This confirms the theory that, if a higher quantile is of
 particular interest, it is beneficial to combine the model information across other quantile levels, for example,
 some moderate quantile $\tau=0.5$, for more efficient and stable estimation.

\begin{table}[!h]
\tabcolsep=1.5pt\fontsize{8}{10}
\selectfont
\begin{center}
{\normalsize Table 2: Simulation results for five models with a high quantile.}\\
\vskip 10pt
\begin{tabular}{c cc cc ccc }
\hline
&& &\multicolumn{2}{c}{$\tau=0.5$} & & \multicolumn{2}{c}{$\tau=0.9$}\\
\cline{4-5} \cline{7-8}
Model&$n$&&$\beta_1(\tau)$&$\beta_2(\tau)$& &$\beta_1(\tau)$&$\beta_2(\tau)$\\
\hline

M1

&& True            & 2 & 1 && 2 & 2.2816 \\
&1000& TQE             &2.0007(0.0512)&0.9974(0.0899)&&2.0117(0.0725)&2.2734(0.1296)\\
&& SEF             &2.0009(0.0238)&0.9968(0.0547)&&2.0158(0.0362)&2.2605(0.0784)\\
&& EFF &2.0014(0.0226)&0.9960(0.0530)&&2.0032(0.0377)&2.2757(0.0772)\\

\cline{3-8}
&2000& TQE             &1.9992(0.0365)&1.0010(0.0652)&&2.0090(0.0482)&2.2727(0.0877)\\
&& SEF             &2.0002(0.0159)&0.9993(0.0361)&&2.0088(0.0231)&2.2698(0.0543)\\
&& EFF &2.0004(0.0142)&0.9989(0.0347)&&2.0026(0.0207)&2.2777(0.0510)\\

\cline{2-8}
M2

&& True            & 2 & 2 & &3.2816 & 3.2816 \\
&1000& TQE             &1.9976(0.1192)&1.9987(0.1155)&&3.2885(0.1622)&3.2751(0.1648)\\
&& SEF             &1.9989(0.0896)&1.9981(0.0875)&&3.2818(0.1227)&3.2764(0.1237)\\
&& EFF &1.9982(0.0879)&1.9984(0.0868)&&3.2839(0.1189)&3.2785(0.1200)\\
\cline{3-8}

&2000& TQE             &1.9980(0.0834)&2.0022(0.0844)&&3.2861(0.1149)&3.2750(0.1127)\\
&& SEF             &1.9990(0.0617)&2.0003(0.0614)&&3.2836(0.0851)&3.2784(0.0862)\\
&& EFF &1.9986(0.0607)&2.0004(0.0607)&&3.2845(0.0821)&3.2797(0.0839)\\

\cline{2-8}
M3

&& True            & 2 & 1 && 2 & 3.1972 \\
&1000& TQE             &2.0011(0.0822)&0.9958(0.1437)&&2.0246(0.1410)&3.1834(0.2531)\\
&& SEF             &2.0014(0.0381)&0.9949(0.0874)&&2.0354(0.0720)&3.1530(0.1533)\\
&& EFF &2.0023(0.0366)&0.9935(0.0848)&&2.0107(0.0668)&3.1817(0.1458)\\

\cline{3-8}
&2000& TQE             &1.9987(0.0585)&1.0017(0.1042)&&2.0186(0.0938)&3.1807(0.1708)\\
&& SEF             &2.0003(0.0256)&0.9990(0.0575)&&2.0201(0.0458)&3.1719(0.1061)\\
&& EFF &2.0005(0.0226)&0.9985(0.0555)&&2.0074(0.0403)&3.1872(0.0995)\\

\cline{2-8}
M4

&& True            & 2 & 1 & &2 & 4.0777 \\
&1000& TQE             &2.0009(0.0669)&0.9966(0.1144)&&2.1044(0.4091)&4.0791(0.7658)\\
&& SEF             &2.0014(0.0316)&0.9955(0.0699)&&2.1874(0.2729)&3.9097(0.4643)\\
&& EFF &2.0023(0.0286)&0.9943(0.0672)&&2.0994(0.2469)&3.9866(0.4447)\\
\cline{3-8}

&2000& TQE             &1.999(0.0469)&1.0013(0.0824)&&2.0754(0.2667)&4.0444(0.5028)\\
&& SEF             &2.0002(0.0207)&0.9993(0.0461)&&2.1081(0.1550)&3.9713(0.3199)\\
&& EFF &2.0007(0.0183)&0.9984(0.0444)&&2.0579(0.1358)&4.0204(0.3075)\\

\cline{2-8}
M5

&& True            & 1 & 2 && 3.1972 & 5.0777 \\
&1000& TQE             &0.9964(0.1797)&1.9982(0.1555)&&3.2258(0.4602)&5.1270(0.8171)\\
&& SEF             &0.9979(0.1344)&1.9972(0.1179)&&3.2108(0.3645)&5.1003(0.6241)\\
&& EFF &0.9968(0.1313)&1.9987(0.1166)&&3.2081(0.3424)&5.1070(0.5705)\\
\cline{3-8}
&2000& TQE             &0.9973(0.1258)&2.0030(0.1139)&&3.2158(0.3218)&5.0801(0.5341)\\
&& SEF             &0.9987(0.0921)&2.0006(0.0831)&&3.2074(0.2518)&5.0805(0.4226)\\
&& EFF &0.9981(0.0906)&2.0005(0.0814)&&3.2088(0.2400)&5.0860(0.3935)\\

\hline
\multicolumn{8}{l}{$^*$ Standard deviations are in parentheses. }
\end{tabular}
\end{center}
\end{table}

\vspace{0.1in}

\vspace{3ex}

\begin{center}
\textsf{4. \hspace{0.1in} APPLICATION}
\end{center}
\vspace{0.1in}
We apply the proposed method to analyze a birth data (birth) released annually by the National Center for Health Statistics.
The data includes information on nearly all live births from United States.
Education of mother of each birth is recorded as 5 classes based on years of education.
For illustration, we only consider the births that occurred in the month of June, 1997, and had mothers with smoking cigarettes and education class 2 (7 to 11 years of education).
There are 9832 birth children consisting of 4861 female and 4971 male. In this paper, our interest is to study the relationship of the birth weight of child (in grams) and the covariates: the age of mother (Mage), the age of father (Fage) and the total number of prenatal care visits (Nprevist). All variables are
taken the logarithmic transformation before analysis.
We apply model (\ref{qrm}) with $\tau=0.3, 0.5, 0.7$ for analyzing the dataset.  Tables 3-4 present the estimation results of regression coeffecients by TQE, SEF and EFF, which are defined the same as in section 3.
In Tables 3-4, Est represents the parameter estimate, Esd is the variance estimate of Est by
$1000$ boostrap resampling method and the $P$-value is computed by $1-\Phi(|Est/Esd|)$ where
$\Phi(\cdot)$ is the cumulative distribution function of the standard normal distribution.

It can be seen that at nominal significance level 0.05, all the three methods
detect Nprevist for all quantiles, detect ages of parents at $\tau=0.3$ and 0.5. And at $\tau=0.7$,
 the three methods identify father age of the female children data.
  However, one significant finding in the analysis is that at $\tau=0.7$, Fage and Mage of the male children data do not have significantly nonzero coefficients, however, for female data, Mage is only detected by EFF with a significant nonzero coefficients, while TQE and SEF do not detect this.
Overall, Tables 3-4 report that Nprevist and ages of parents have positive and negative coefficients, respectively, which suggests that
the birth weights of children become heavier when their mothers are younger and have more prenatal care visits.  In addition, the
effect of the three covariates to the birth weights of children are more significant at lower quantile ($\tau=0.3$) compared with that of higher quantile ($\tau=0.7$).

\begin{table}[!h]
\tabcolsep=3pt\fontsize{7}{9}
\selectfont
\begin{center}
{\normalsize Table 3: Analysis of birth data with male child.}\\
\vskip 10pt
\begin{tabular}{ccccc ccccc ccccc cc}
\hline
&&\multicolumn{3}{c}{Intercept} & & \multicolumn{3}{c}{Mage}& & \multicolumn{3}{c}{Fage}& & \multicolumn{3}{c}{Nprevist}\\
\cline{3-5} \cline{7-9} \cline{11-13} \cline{15-17}
$\tau$&model&Est& Esd &P value && Est& Esd &P value &&   Est& Esd &P value &&   Est& Esd &P value\\
\hline

0.3&TQE&8.1087&0.0536&$<0.0001$&&-0.0540&0.0161&0.0004&&-0.0144&0.0054&0.0038&&0.0388&0.0049&$<0.0001$\\
&SEF&8.1009&0.0544&$<0.0001$&&-0.0523&0.0164&0.0007&&-0.0135&0.0054&0.0067&&0.0384&0.0047&$<0.0001$\\
&EFF&8.0972&0.0608&$<0.0001$&&-0.0510&0.0182&0.0026&&-0.0146&0.0059&0.0069&&0.0397&0.0061&$<0.0001$\\
0.5&TQE&8.1311&0.0475&$<0.0001$&&-0.0362&0.0141&0.0051&&-0.0096&0.0046&0.0177&&0.0357&0.0047&$<0.0001$\\
&SEF&8.1137&0.0483&$<0.0001$&&-0.0313&0.0148&0.0168&&-0.0092&0.0048&0.0269&&0.0359&0.0052&$<0.0001$\\
&EFF&8.1233&0.0545&$<0.0001$&&-0.0328&0.0158&0.0186&&-0.0096&0.0049&0.0245&&0.0345&0.0051&$<0.0001$\\
0.7&TQE&8.1217&0.0415&$<0.0001$&&-0.0154&0.0122&0.1029&&-0.0039&0.0039&0.1597&&0.0357&0.0045&$<0.0001$\\
&SEF&8.1197&0.0427&$<0.0001$&&-0.0146&0.0125&0.1211&&-0.0040&0.0039&0.1531&&0.0357&0.0046&$<0.0001$\\
&EFF&8.1217&0.0448&$<0.0001$&&-0.0147&0.0133&0.1338&&-0.0041&0.0042&0.1640&&0.0352&0.0049&$<0.0001$\\

\hline
\end{tabular}
\end{center}
\end{table}

\begin{table}[!h]
\tabcolsep=3pt\fontsize{7}{9}
\selectfont
\begin{center}
{\normalsize Table 4: Analysis of birth data with female child.}\\
\vskip 10pt
\begin{tabular}{ccccc ccccc ccccc cc}
\hline
&&\multicolumn{3}{c}{Intercept} & & \multicolumn{3}{c}{Mage}& & \multicolumn{3}{c}{Fage}& & \multicolumn{3}{c}{Nprevist}\\
\cline{3-5} \cline{7-9} \cline{11-13} \cline{15-17}
$\tau$&model&Est& Esd &P value && Est& Esd &P value &&   Est& Esd &P value &&   Est& Esd &P value\\
\hline

0.3&TQE&8.2687&0.0548&$<0.0001$&&-0.1058&0.0177&$<0.0001$&&-0.0172&0.0049&0.0002&&0.0258&0.0044&$<0.0001$\\
&SEF&8.2254&0.0533&$<0.0001$&&-0.0947&0.0174&$<0.0001$&&-0.0143&0.0052&0.0030&&0.0250&0.0049&$<0.0001$\\
&EFF&8.3082&0.1003&$<0.0001$&&-0.1152&0.0263&$<0.0001$&&-0.0207&0.0081&0.0056&&0.0271&0.0114&0.0087\\
0.5&TQE&8.1886&0.0455&$<0.0001$&&-0.0498&0.0126&$<0.0001$&&-0.0156&0.0048&0.0006&&0.0210&0.0051&$<0.0001$\\
&SEF&8.1815&0.0440&$<0.0001$&&-0.0478&0.0124&0.0001   &&-0.0153&0.0049&0.0010&&0.0209&0.0051&$<0.0001$\\
&EFF&8.1829&0.0511&$<0.0001$&&-0.0484&0.0140&0.0003   &&-0.0151&0.0050&0.0013&&0.0209&0.0055&0.0001\\
0.7&TQE&8.1628&0.0407&$<0.0001$&&-0.0175&0.0130&0.0902   &&-0.0151&0.0042&0.0001&&0.0212&0.0044&$<0.0001$\\
&SEF&8.1497&0.0412&$<0.0001$&&-0.0119&0.0133&0.1860   &&-0.0155&0.0042&0.0001&&0.0199&0.0045&$<0.0001$\\
&EFF&8.1925&0.0513&$<0.0001$&&-0.0258&0.0152&0.0450   &&-0.0150&0.0049&0.0011&&0.0200&0.0053&0.0001\\

\hline
\end{tabular}
\end{center}
\end{table}

%

\newpage

\begin{center}
\textsf{ \hspace{0.1in} APPENDIX }
\end{center}
\vspace{0.1in}
\setcounter{equation}{0}
 \renewcommand\theequation{A.\arabic {equation} }

\noindent \textbf{Appendix I}

We firstly consider the following quantile regression model,
\begin{eqnarray}
\label{1-1}Q_{Y|X}(\tau_l)=X^\top{\bm \beta}(\tau_l), ~~~~
~~l=1,2,\cdots,L,\nonumber
\end{eqnarray}
where $X=(X_1,X_2,\cdots,X_p)^\top$, ${\bm \beta}(t)=(\beta_1(t),\beta_2(t),\cdots,\beta_p(t))^\top$. This model focuses on parameter estimation of the $L$ quantile points.

A semiparametric efficient score of ${\bm{\beta}}(\tau_k)$ is calculated by using methodology of the least favorable submodel (Bickel {\it et al.}, 1993). Without loss of generality, the efficient score of the $j$-th component of ${\bm{\beta}}(\tau_k)$  is constructed firstly. We consider the following parametric submodels based on cumulative distribution function with parameter $\theta$ in a neighborhood of 0,
\begin{eqnarray}
\tilde F_{Y|X}(t;\theta)=F_{Y|X}(t)+\theta G_{Y|X}(t),\label{sub-model}
\end{eqnarray}
where $G_{Y|X}(t)$ is a function on $t$. Easily, we have
\begin{eqnarray}
\tilde f_{Y|X}(t;\theta)=f_{Y|X}(t)+\theta g_{Y|X}(t),\label{sub-model1}
\end{eqnarray}
where $\tilde f_{Y|X}(t;\theta)$, $f_{Y|X}(t)$ and $g_{Y|X}(t)$ are derivatives of $\tilde F_{Y|X}(t;\theta)$, $F_{Y|X}(t)$
and $G_{Y|X}(t)$ with respect to $t$.

To guarantee the $\tilde f_{Y|X}(t;\theta)$ is a density function
for all of $\theta$, the $g_{Y|X}(t)$ satisfies that
\begin{eqnarray}\label{eq1-2}
\int_{-\infty}^{+\infty} g_{Y|X}(u) du=0.
\end{eqnarray}
Let $x^\top{\bm{\beta}}(\tau_l;\theta)$ be the $\tau_l$ quantile of distribution of $\tilde F_{Y|X}(t;\theta)$ and
$x^\top{\bm{\beta}}(\tau_l;0)=x^\top{\bm{\beta}}_{0}(\tau_l)$,  for $l=1,2,\ldots,L$. From (\ref{sub-model}) and the Taylor expansion, we have
\begin{align}
\tau_l=&\tilde{F}_{Y|X}(x^\top{\bm{\beta}}(\tau_l;\theta);\theta)\nonumber\\
=&F_{Y|X}(x^\top{\bm{\beta}}_{0}(\tau_l))+f_{Y|X}(x^\top{\bm{\beta}}_{0}(\tau_l))
x^\top({\bm{\beta}}(\tau_l;\theta)-{\bm \beta}_{0}(\tau_l))+\theta G_{Y|X}(x^\top{\bm{\beta}}_{0}(\tau_l))
+o(|\theta|)\nonumber\\
=&\tau_l+f_{Y|X}(x^\top{\bm{\beta}}_{0}(\tau_l))
x^\top({\bm{\beta}}(\tau_l;\theta)-{\bm \beta}_{0}(\tau_l))+ \theta G_{Y|X}(x^\top{\bm{\beta}}_{0}(\tau_l))+
o(|\theta|),\nonumber
\end{align}
which suggests that
\begin{eqnarray}\label{eq1-3}
G_{Y|X}(x^\top{\bm \beta}_{0}(\tau_l))\theta=-f_{Y|X}(x^\top{\bm \beta}_{0}(\tau_l))
({\bm{\beta}}(\tau_l;\theta)-{\bm \beta}_{0}(\tau_l))^\top x+o(|\theta|).
\end{eqnarray}
Let ${\bm d}(\tau_l)$ be derivative value of ${\bm{\beta}}(\tau_l;\theta)$ on $\theta=0$. Note that ${\bm d}(\tau_l)$ is a vector with length $p$ and
${\bm{\beta}}(\tau_l;\theta)-{\bm \beta}_{l0}={\bm d}(\tau_l)\theta+o(|\theta|)$.
From (\ref{eq1-3}), we have
\begin{eqnarray}\label{eq1-3-1}
G_{Y|X}(x^\top{\bm \beta}_{0}(\tau_l))\theta=-f_{Y|X}(x^\top{\bm \beta}_{0}(\tau_l))
x^\top{\bm d}(\tau_l)\theta+o(|\theta|),
\end{eqnarray}
which indicates that
\begin{eqnarray}\label{eq1-4}
G_{Y|X}(x^\top{\bm \beta}_{0}(\tau_l))=-f_{Y|X}(x^\top{\bm \beta}_{0}(\tau_l))x^\top{\bm d}(\tau_l),~~\mbox{for}~~l=1,2
\ldots,L.
\end{eqnarray}

Thus, we study the studied parametric submodel (\ref{sub-model}), subject to constraints (\ref{eq1-2}) and (\ref{eq1-4}).
It is well-known that $\mbox{Var}(\beta_{j}(\tau_k;\hat\theta))$ can be generally expressed with $d_{j}(\tau_k)^2\mbox{Var}(\hat\theta)$, where $d_{j}(\tau_k)$ and $\beta_{j}(\tau_k;\hat\theta)$ are the $j-$th components of
${\bm d}(\tau_k)$ and ${\bm \beta}_k$ respectively, and $\hat\theta$ is an estimator of $\theta$. When $d_{j}(\tau_k)$ equals to 1, $\mbox{Var}(\beta_{j}(\tau_k;\hat\theta))$ can be approximated
 by $\mbox{Var}(\hat\theta)$. This paper takes $d_{j}(\tau_k)=1$.

Based on the density function $\tilde f_{Y|X}(t)$, we show
\begin{eqnarray}\label{eq1-5}
\frac{\partial \log \tilde f_{Y|X}(t)}{\partial \theta}\bigg |_{\theta=0}=\frac{g_{Y|X}(t)}{f_{Y|X}(t)}\doteq \xi.
\end{eqnarray}
Then we have
\begin{align}\label{eq1-6}
E(\xi^2|x)&=\left(\int_{-\infty}^{x^\top{\bm \beta}_{0}(\tau_1)}+\sum_{l=2}^L\int_{x^\top{\bm \beta}_{0}(\tau_{l-1})}
^{x^\top{\bm \beta}_{0}(\tau_l)}+\int_{x^\top{\bm \beta}_{0}(\tau_L)}^{+\infty}\right)
\frac{g^2_{Y|X}(t)}{f_{Y|X}(t)}dt\nonumber\\
&=\left(\int_{-\infty}^{x^\top{\bm \beta}_{0}(\tau_1)}+\sum_{l=2}^L\int_{x^\top{\bm \beta}_{0}(\tau_{l-1})}
^{x^\top{\bm \beta}_{0}(\tau_l)}+\int_{x^\top{\bm \beta}_{0}(\tau_L)}^{+\infty}\right)
\Big(\frac{g_{Y|X}(t)}{f^{1/2}_{Y|X}(t)}\Big)^2dt\nonumber\\
&\geq\frac{(\int_{-\infty}^{x^\top{\bm \beta}_{0}(\tau_1)}{g_{Y|X}(t)}dt)^2}{
\int_{-\infty}^{x^\top{\bm \beta}_{0}(\tau_1)} {f_{Y|X}(t)}dt}+\sum_{l=2}^L
\frac{(\int_{x^\top{\bm \beta}_{0}(\tau_{l-1})}^{x^\top{\bm \beta}_{0}(\tau_l)}{g_{Y|X}(t)}dt)^2}{
\int_{x^\top{\bm \beta}_{0}(\tau_{l-1})}^{x^\top{\bm \beta}_{0}(\tau_l)} {f_{Y|X}(t)}dt}+
\frac{(\int_{x^\top{\bm \beta}_{0}(\tau_L)}^{+\infty}{g_{Y|X}(t)}dt)^2}{
\int_{x^\top{\bm \beta}_{0}(\tau_L)}^{+\infty} {f_{Y|X}(t)}dt}\nonumber\\
&=\sum_{l=1}^{L+1}\frac{(G(x^\top{\bm \beta}_{0}(\tau_l))-G(x^\top{\bm \beta}_{0}(\tau_{l-1})))^2}
{\tau_l-\tau_{l-1}}\nonumber\\
&=\sum_{l=1}^{L+1}\frac{(f_{Y|X}(x^\top{\bm \beta}_{0}(\tau_{l-1}))x^\top{\bm d}(\tau_{l-1})
-f_{Y|X}(x^\top{\bm \beta}_{0}(\tau_l))x^\top{\bm d}(\tau_l))^2}{\tau_l-\tau_{l-1}},
\end{align}
where ${\bm d}(\tau_0)={\bm d}(\tau_{L+1})=0$.
Hence, we get
\begin{align}\label{eq1-6-1}
E(\xi^2)=E(E(\xi^2|x))\geq&\sum_{l=1}^{L+1}\frac{E\Big(f_{Y|X}(x^\top{\bm \beta}_{0}(\tau_{l-1}))x^\top{\bm d}(\tau_{l-1})
-f_{Y|X}(x^\top{\bm \beta}_{0}(\tau_l))x^\top{\bm d}(\tau_l)\Big)^2}{\tau_l-\tau_{l-1}}\nonumber\\
=&{\bm d}^\top {\bm B} {\bm A} {\bm B}^\top {\bm d},
\end{align}
where ${\bm d}=({\bm d}(\tau_1)^\top,{\bm d}(\tau_2)^\top,\ldots,{\bm d}(\tau_L)^\top)^{T}$,
\begin{align}
&{\bm A}_1=\left(\frac{E(f_{Y|X}(X^\top{\bm \beta}_0(\tau_1))^2XX^\top)}{\tau_1}\right)_{p\times p},~~{\bm A}_{L+1}=\left(\frac{E(f_{Y|X}(X^\top{\bm \beta}_0(\tau_L))^2XX^\top)}{1-\tau_L}\right)_{p\times p},\nonumber\\
&{\bm A}_l=\left(\begin{array}{cc}
  \frac{E(f_{Y|X}(X^\top{\bm \beta}_0(\tau_{l-1}))^2XX^\top)}{\tau_l-\tau_{l-1}} &  -\frac{E(f_{Y|X}(X^\top{\bm \beta}_0(\tau_{l-1}))f_{Y|X}(X^\top{\bm \beta}_0(\tau_l))XX^\top)}{\tau_l-\tau_{l-1}} \\
 -\frac{E(f_{Y|X}(X^\top{\bm \beta}_0(\tau_{l-1}))f_{Y|X}(X^\top{\bm \beta}_0(\tau_l))XX^\top)}{\tau_l-\tau_{l-1}}  &  \frac{E(f_{Y|X}(X^\top{\bm \beta}_0(\tau_l))^2XX^\top)}{\tau_l-\tau_{l-1}}
\end{array}
\right)_{2p\times 2p},\nonumber\\
&~~~~l=2,\cdots,L\nonumber
\end{align}
\begin{align}
{\bm A}=\left(\begin{array}{ccccc}
  {\bm A}_1 &  0& ... & ... & 0 \\
  0 & {\bm A}_2 &  0 &  ... & 0 \\
  ... & ... & ... &...  &  ...\\
  ... & 0 & ... & {\bm A}_L & 0  \\
  0 & ... & ... & 0 &  {\bm A}_{L+1}
\end{array}
\right)_{2pL\times 2pL},\label{Adef}
\end{align}
\begin{align}
{\bm B}=\left(\begin{array}{cccccccccc}
  I_p &  I_p & 0 & 0& ... & &  & ... & 0 & 0 \\
  0 & 0 &  I_p &  I_p & 0 & ... & ... & ... & 0 & 0 \\
   &  & ... & ... &  &  &...  &...  &  &  \\
  0 & 0& ... & & ...& 0 &  I_p &  I_p& 0 & 0 \\
  0 & 0 & ... &  &  & ... & 0 & 0 &  I_p &  I_p
\end{array}
\right)_{pL\times 2pL}.\label{Bdef}
\end{align}

The equation holds if and only if $$\frac{g_{Y|X}(t)}{f^{1/2}_{Y|X}(t)}=
\sum_{l=1}^{L+1}\{a_lf^{1/2}_{Y|X}(t)+b_l\}I(x^\top{\bm \beta}_0(\tau_{l-1})<y<x^\top
{\bm \beta}_0(\tau_l)).$$
It follows from conditions (\ref{eq1-2}) and (\ref{eq1-4}) that
$$b_l=0,~~a_l=\frac{\Big(f_{Y|X}(x^\top{\bm \beta}_0(\tau_{l-1}))
{\bm d}(\tau_{l-1})-f_{Y|X}(x^\top{\bm \beta}_0(\tau_{l})){\bm d}(\tau_{l})\Big)^\top x}{\tau_l-\tau_{l-1}},~~l=1,\cdots,L+1.$$
Therefore, a semiparametric efficient score
for ${\bm \beta}_{j}(\tau_k)$ is,
\begin{align}
&\frac{g_{Y|X}(t)}{f_{Y|X}(t)}\nonumber\\
=&\sum_{l=1}^{L+1}\frac{f_{Y|X}(x^\top{\bm \beta}_0(\tau_{l-1}))
x^\top{\bm d}(\tau_{l-1})-f_{Y|X}(x^\top{\bm \beta}_0(\tau_{l}))
x^\top{\bm d}(\tau_{l})}{\tau_l-\tau_{l-1}}\nonumber\\
&I(x^\top{\bm \beta}_0(\tau_{l-1})<y<x^\top{\bm \beta}_0(\tau_{l}))\nonumber\\
=&\sum_{l=1}^{L+1}\frac{f_{Y|X}(x^\top{\bm \beta}_0(\tau_{l-1}))
x^\top{\bm d}(\tau_{l-1})-f_{Y|X}(x^\top{\bm \beta}_0(\tau_{l}))
x^\top{\bm d}(\tau_{l})}{\tau_l-\tau_{l-1}}\nonumber\\
&\Big\{I(x^\top{\bm \beta}_0(\tau_{l-1})<y<x^\top{\bm \beta}_0(\tau_{l}))-(\tau_l-\tau_{l-1})\Big\}.\label{eq1-7}
\end{align}

Due to unknown ${\bm d}$, we need to compute ${\bm d}$ by minimizing
$E(\xi^2)$, that is minimizing quadratic function ${\bm d}^\top {\bm B} {\bm A} {\bm B}^\top {\bm d}$ on ${\bm d}$ subject to $d_{j}(\tau_k)=1$.
Let ${\bm U}={\bm B} {\bm A} {\bm B}^\top$ and ${\bm W}$ be a $p\times p$ diagonal matrix with diagonal components same as $1/{\mbox {diag}}({\bm U}^{-1})$.
Denote $({\bm u}_1,{\bm u}_2,\cdots,{\bm u}_{pL})={\bm U}^{-1}{\bm W}$. By Lagrange multiplier method, we have
\begin{eqnarray}
L({\bm d},\lambda)={\bm d}^\top{\bm U}{\bm d}+\lambda\{d_{j}(\tau_k)-1\}.\nonumber
\end{eqnarray}
From ${\partial L({\bm d},\lambda)}/{\partial {\bm d}}=0$ and $d_{j}(\tau_k)=1$, we get
\begin{eqnarray}\label{req-1}
{\bm d}={\bm u}_{kj}.
\end{eqnarray}
Therefore, for $1\leq j\leq p$, the semiparametric efficient score of ${\bm{\beta}}_{j}(\tau_k)$ is
\begin{align}
S_{kj}(y,x)=&\sum_{l=1}^{L+1}\frac{f_{Y|X}(x^\top{\bm \beta}_0(\tau_{l-1}))
x^\top{\bm d}(\tau_{l-1})-f_{Y|X}(x^\top{\bm \beta}_0(\tau_{l}))
x^\top{\bm d}(\tau_{l})}{\tau_l-\tau_{l-1}}\nonumber\\
&\Big\{I(x^\top{\bm \beta}_0(\tau_{l-1})<y<x^\top{\bm \beta}_0(\tau_{l}))
-(\tau_l-\tau_{l-1})\Big\},\nonumber
\end{align}
with ${\bm d}={\bm u}_{kj}$. Naturally,
the semiparametric efficient score of ${\bm{\beta}}(\tau_k)$ is
\begin{align}
S_{k}(y,x)=&\sum_{l=1}^{L+1}\frac{f_{Y|X}(x^\top{\bm \beta}_0(\tau_{l-1}))
x^\top{\bm D}_{l-1}-f_{Y|X}(x^\top{\bm \beta}_0(\tau_{l}))
x^\top{\bm D}_{l}}{\tau_l-\tau_{l-1}}\nonumber\\
&\Big\{I(x^\top{\bm \beta}_0(\tau_{l-1})<y<x^\top{\bm \beta}_0(\tau_{l}))
-(\tau_l-\tau_{l-1})\Big\},\nonumber
\end{align}
where $({\bm D}_1,{\bm D}_2,\ldots,{\bm D}_L)=({\bm u}_{k1},{\bm u}_{k2},\ldots,{\bm u}_{kp})^\top$, ${\bm D}_l$ is $p\times p$ matrix, $l=0,1,\ldots,L+1$, and ${\bm D}_0={\bm D}_{L+1}=0$.

\vskip 0.5cm

\noindent \textbf{Appendix II}\\

Consider another quantile regression model,
\begin{eqnarray}
Q_{Y|X}(\tau)=x^\top{\bm \beta}(\tau), ~~~~
~~0<\tau<1,\nonumber
\end{eqnarray}
where $x=(x_1,x_2,\cdots,x_p)^\top$, ${\bm \beta}(t)=(\beta_1(t),\beta_2(t),\cdots,\beta_p(t))^\top$.
This model assumes all of quantiles for response $Y$ given $X$ have a linear form. Since the cumulative distribution function of $Y$ given $X$ is $F_{Y|X}({\bm \beta}(\tau)^\top X)=\tau$, and density function of $Y$ given $X=x$, satisfies $$f_{Y|X}({\bm \beta}(\tau)^\top x)=\frac{1}{x^\top\dot{\bm \beta}(\tau)}. $$

 A semiparametric efficient score of ${\bm{\beta}}_j(\tau_k),~~j=1,2,\ldots,p,$ is calculated by using the least favorable submodel similar to the previous parametric submodel. The parametric submodels with parameter $\theta$ in a neighborhood of 0 is,
\begin{eqnarray}
\tilde F_{Y|X}(t;\theta)=F_{Y|X}(t)+\theta G_{Y|X}(t),\label{sub-model2-1}
\end{eqnarray}
where $G_{Y|X}(t)$ is a function on $t$. Then we have
\begin{eqnarray}
\tilde f_{Y|X}(t;\theta)=f_{Y|X}(t)+\theta g_{Y|X}(t),\label{sub-model2-2}
\end{eqnarray}
where $\tilde f_{Y|X}(t;\theta)$, $f_{Y|X}(t)$ and $g_{Y|X}(t)$ are derivatives of $\tilde F_{Y|X}(t;\theta)$, $F_{Y|X}(t)$
and $G_{Y|X}(t)$ with respect to $t$. Similar derivations to (\ref{eq1-2}) and (\ref{eq1-4}), we have constraints on $g$ as follows,
\begin{eqnarray}\label{eq2-2}
\int_{-\infty}^{+\infty} g_{Y|X}(u) du=0,
\end{eqnarray}
and
\begin{eqnarray}\label{eq2-4}
G_{Y|X}(x^\top{\bm \beta}_{0}(\tau))=-f_{Y|X}(x^\top{\bm \beta}_{0}(\tau))x^\top{\bm d}(\tau),~~\mbox{for}~~0<\tau<1,
\end{eqnarray}
where ${\bm d}(\tau)$ is derivative value of ${\bm{\beta}}(\tau;\theta)$ with respect to $\theta$ at point $0$, and ${d}_j(\tau_k)=1$; $x^\top{\bm \beta}_{0}(\tau)$ and $x^\top{\bm \beta}(\tau;\theta)$ are $\tau$ quantiles of $F_{Y|X}(t)$ and $\tilde F_{Y|X}(t;\theta)$, respectively; and ${\bm \beta}(\tau;0)={\bm \beta}_0(\tau)$. Since $x^\top{\bm \beta}_{0}(\tau)$ is a monotone and increasing function on $\tau$, from (\ref{eq2-2}) we have $G_{Y|X}(x^\top{\bm \beta}_{0}(0))=G_{Y|X}(x^\top{\bm \beta}_{0}(1))=0$.

From (\ref{eq2-4}), it shows
$$\frac{\partial G_{Y|X}(x^\top{\bm \beta}_{0}(\tau))}{\partial \tau}=-\frac{\partial (f_{Y|X}(x^\top{\bm \beta}_{0}(\tau))x^\top{\bm d}(\tau))}{\partial \tau},$$
which indicates that
\begin{eqnarray}
\frac{g_{Y|X}(x^\top{\bm \beta}_{0}(\tau))}{f_{Y|X}(x^\top{\bm \beta}_{0}(\tau))}=g_{Y|X}(x^\top{\bm \beta}_{0}(\tau))x^\top\dot{\bm \beta}_0(\tau)=-\frac{\partial (f_{Y|X}(x^\top{\bm \beta}_{0}(\tau))x^\top{\bm d}(\tau))}{\partial \tau}\nonumber\\
\doteq-\dot{\Big(
f_{Y|X}(x^\top{\bm \beta}_{0}(\tau))x^\top{\bm d}(\tau)\Big)}.
\end{eqnarray}
Hence, we have
\begin{eqnarray}\label{eq2-5}
\frac{\partial \log \tilde f_{Y|X}(t)}{\partial \theta}\bigg |_{\theta=0}=\frac{g_{Y|X}(t)}{f_{Y|X}(t)}=
-\dot{\Big(
f_{Y|X}(x^\top{\bm \beta}_{0}(\tau))x^\top{\bm d}(\tau)\Big)}\bigg |_{t=x^\top{\bm \beta}_{0}(\tau)}
\doteq \xi,
\end{eqnarray}
and  the semiparametric efficient score
of ${\bm \beta}_{j}(\tau_k)$,
\begin{align}\label{s2}
S_{kj}(y,x)=&-\dot{\Big(
f_{Y|X}(x^\top{\bm \beta}_{0}(\tau))x^\top{\bm d}(\tau)\Big)},
\end{align}
where ${\bm d}(\tau)$ is a minimizer of
$$E(\xi^2)=\int_{-\infty}^{\infty}\left\{\dot{\Big(
f_{Y|X}(x^\top{\bm \beta}_{0}(\tau))x^\top{\bm d}(\tau)\Big)}\bigg |_{t=x^\top{\bm \beta}_{0}(\tau)}\right\}^2dt,$$
subject to ${d}_j(\tau_k)=1$.

Obviously, it is intractable to compute the semiparametric
score (\ref{s2}) of ${\bm \beta}_{j}(\tau_k)$, and can not be used to estimate the ${\bm\beta}_0(\tau)$ directly.

\vskip 0.5cm

\noindent \textbf{Appendix III}\\

For the $j$th component of ${\bm \beta}(\tau_k)$, $j=1,...,p$,
it follows from
(\ref{m-s1}) that ${\bm d}$ can be solved by minimizing quadratic form
${\bm d}^\top {\bm B} {\bm A} {\bm B}^\top {\bm d}$ on ${\bm d}$, subject to $d_{j}(\tau_k)=1$.
For $l\neq k$, letting derivative of ${\bm d}^\top {\bm B} {\bm A} {\bm B}^\top {\bm d}$ on ${\bm d}(\tau_l)$
be 0, we have
\begin{align}\label{dexp}
&(\tau_{l+1}-\tau_l)E\Big(f_{Y|X}(X^\top{\bm \beta}(\tau_{l-1}))f_{Y|X}(X^\top{\bm \beta}(\tau_{l}))XX^\top\Big){\bm d}(\tau_{l-1})\nonumber\\
&-
(\tau_{l+1}-\tau_{l-1})E\Big(f_{Y|X}(X^\top{\bm \beta}(\tau_{l}))f_{Y|X}(X^\top{\bm\beta}(\tau_{l}))XX^\top\Big){\bm d}(\tau_{l})\nonumber\\
&+
(\tau_{l}-\tau_{l-1})E\Big(f_{Y|X}(X^\top{\bm\beta}(\tau_{l}))f_{Y|X}(X^\top{\bm\beta}(\tau_{l+1})
)XX^\top\Big){\bm d}(\tau_{l+1})=0.
\end{align}
Based on model (\ref{qrm-2}) and cumulative distribution function (\ref{cdf}) with $p=1$, we have
\begin{eqnarray}
f_{Y|X}(X{\bm \beta}(\tau_{l}))=\frac{1}{X\dot{{\bm \beta}}(\tau_l)},~~l=1,2,\ldots,L,\nonumber
\end{eqnarray}
which leads to
\begin{eqnarray}\label{deqn1}
f_{Y|X}(X{\bm\beta}(\tau_{l}))X=\frac{1}{\dot{ {\bm\beta}}(\tau_l)},~~l=1,2,\ldots,L.\nonumber
\end{eqnarray}
Thus, from (\ref{dexp}), we show that
\begin{align}\label{deqn}
0=&(\tau_{l+1}-\tau_l)\frac{{\bm d}(\tau_{l-1})}{\dot{\bm \beta}(\tau_{l-1})}-
(\tau_{l+1}-\tau_{l-1})\frac{{\bm d}(\tau_{l})}{\dot{\bm \beta}(\tau_{l})}+
(\tau_{l}-\tau_{l-1})\frac{{\bm d}(\tau_{l+1})}{\dot{\bm \beta}(\tau_{l+1})}\nonumber\\
=&(\tau_{l+1}-\tau_l)\Big(\frac{{\bm d}(\tau_{l-1})}{\dot{\bm \beta}(\tau_{l-1})}-\frac{{\bm d}(\tau_{l})}{\dot{\bm \beta}(\tau_{l})}\Big)-
(\tau_{l}-\tau_{l-1})\Big(\frac{{\bm d}(\tau_{l})}{\dot{\bm \beta}(\tau_{l})}-\frac{{\bm d}(\tau_{l+1})}{\dot{\bm \beta}(\tau_{l+1})}\Big).
\end{align}
Hence, the score (\ref{m-s1}) becomes
\begin{align}\label{m-s3}
&S_{kj}(y,x)\nonumber\\
=&\sum_{l=1}^{L+1}\frac{f_{Y|X}(x{\bm \beta}(\tau_{l-1}))
x{\bm d}(\tau_{l-1})-f_{Y|X}(x{\bm \beta}(\tau_{l}))
x{\bm d}(\tau_{l})}{\tau_l-\tau_{l-1}}I(x{\bm \beta}(\tau_{l-1})<y<x{\bm \beta}(\tau_{l}))\nonumber\\
=&\sum_{l=1}^{L}\Big(\frac{f_{Y|X}(x{\bm \beta}(\tau_{l}))x{\bm d}(\tau_{l})-f_{Y|X}(x{\bm \beta}(\tau_{l+1}))x{\bm d}(\tau_{l+1})}{\tau_{l+1}-\tau_{l}}\nonumber\\
&\hskip 1cm-\frac{f_{Y|X}(x{\bm \beta}(\tau_{l-1}))x{\bm d}(\tau_{l-1})-f_{Y|X}(x{\bm \beta}(\tau_{l}))x{\bm d}(\tau_{l})}{\tau_l-\tau_{l-1}}\Big)(\tau_l-I(y<{\bm\beta}(\tau_{l})x))\nonumber\\
=&\sum_{l=1}^{L}\left(\frac{{{\bm d}(\tau_{l})}/{\dot{\bm \beta}(\tau_{l})}-{{\bm d}(\tau_{l+1})}/{\dot{\bm \beta}(\tau_{l+1})}}{\tau_{l+1}-\tau_{l}}-\frac{{{\bm d}(\tau_{l-1})}/{\dot{\bm \beta}(\tau_{l-1})}-{{\bm d}(\tau_{l})}/{\dot{\bm \beta}(\tau_{l})}}{\tau_l-\tau_{l-1}}\right)\nonumber\\
&(\tau_l-I(y<{\bm\beta}(\tau_{l})x))\nonumber\\
=&\left(\frac{{{\bm d}(\tau_{k})}/{\dot{\bm \beta}(\tau_{k})}-{{\bm d}(\tau_{k+1})}/{\dot{\bm \beta}(\tau_{k+1})}}{\tau_{k+1}-\tau_{k}}-\frac{{{\bm d}(\tau_{k-1})}/{\dot{\bm \beta}(\tau_{k-1})}-{{\bm d}(\tau_{k})}/{\dot{\bm \beta}(\tau_{k})}}{\tau_k-\tau_{k-1}}\right)\nonumber\\
&(\tau_k-I(y<{\bm \beta}(\tau_{k})x)).
\end{align}

\vskip 0.5cm

\noindent \textbf{Appendix IV}\\

We take estimator of $\beta_1(\tau_1)$ as an example with $p\geq 2$ and $L=2$.
If only use single quantile $\tau_1$ without considering model information of quantile $\tau_2$, from (\ref{m-s1}) with $L=1$, we have
\begin{eqnarray}\label{sigtauvar}
E(S_{11}(Y,X)^2)=E(\frac{1}{\tau_1(1-\tau_1)}f_{Y|X}(X^\top{\bm\beta}(\tau_1))
^2(X^\top{\bm d}(\tau_1))^2)\doteq E(Q_1).
\end{eqnarray}
And taking quantile $\tau_2$ into account ($L=2$ in (\ref{m-s1})), we get
\begin{align}\label{twotauvar}
&E(S_{11}(Y,X)^2)=E\Big\{\frac{1}{\tau_1}f_{Y|X}(X^\top{\bm\beta}(\tau_1))^2
(X^\top{\bm d}(\tau_1))^2+\frac{1}{1-\tau_2}f_{Y|X}(X^\top{\bm\beta}(\tau_2))^2(X^\top{\bm d}(\tau_2))^2\nonumber\\
&+
\frac{1}{\tau_2-\tau_1}\Big(f_{Y|X}(X^\top{\bm \beta}(\tau_1))X^\top{\bm d}(\tau_1)-
f_{Y|X}(X^\top{\bm\beta}(\tau_2))X^\top{\bm d}(\tau_2)\Big)^2\Big\}\doteq E(Q_2).
\end{align}
Then we have
\begin{align}\label{two-one}
&Q_2-Q_1\nonumber\\
=&\frac{1}{\tau_1}(f_{Y|X}(X^\top{\bm\beta}(\tau_1)))^2
(X^\top{\bm d}(\tau_1))^2+\frac{1}{1-\tau_2}(f_{Y|X}(X^\top{\bm\beta}(\tau_2)))^2(X^\top{\bm d}(\tau_2))^2\nonumber\\
&+
\frac{1}{\tau_2-\tau_1}\Big(f_{Y|X}(X^\top{\bm \beta}(\tau_1))X^\top{\bm d}(\tau_1)-
f_{Y|X}(X^\top{\bm\beta}(\tau_2))X^\top{\bm d}(\tau_2)\Big)^2\nonumber\\
&-
\frac{1}{\tau_1(1-\tau_1)}f_{Y|X}(X^\top{\bm\beta}(\tau_1))
^2(X^\top{\bm d}(\tau_1))^2\nonumber\\
=&\frac{1}{\tau_2-\tau_1}\Big(\frac{1-\tau_2}{1-\tau_1}f_{Y|X}(X^\top{\bm\beta}(\tau_1))
^2(X^\top{\bm d}(\tau_1))^2+\frac{1-\tau_1}{1-\tau_2}f_{Y|X}(X^\top{\bm\beta}(\tau_2))
^2(X^\top{\bm d}(\tau_2))^2\Big)\nonumber\\
&-\frac{2}{\tau_2-\tau_1}
f_{Y|X}(X^\top{\bm\beta}(\tau_1))(X^\top{\bm d}(\tau_1))
f_{Y|X}(X^\top{\bm\beta}(\tau_2))(X^\top{\bm d}(\tau_2))\nonumber\\
\geq& 0.
\end{align}
We know that when $Q_2-Q_1=0$ holds, there exist two constants $a$ and $b$ for all $X=x$ such that
$$\sqrt{\frac{1-\tau_2}{1-\tau_1}}f_{Y|X}(x^\top{\bm\beta}(\tau_1))(x^\top{\bm d}(\tau_1))=a\sqrt{\frac{1-\tau_1}{1-\tau_2}}f_{Y|X}(x^\top{\bm\beta}(\tau_2))(x^\top{\bm d}(\tau_2))+b,$$
which obviously does not satisfy.

\vspace{0.2in}
\noindent \textbf{Appendix V: Proofs of Theorem 1 on the one-step efficient estimation. }

\vspace{0.2in}

Let $\hat {\bm \beta}^c(\tau)$ be the classical Koenker-Bassett regression quantile estimator of ${\bm \beta}_0(\tau)$ at any given quantile level $\tau$ and let $h$ be the bandwidth
for the estimation of $\dot{{\bm \beta}_0}(\tau)$, the derivative of ${\bm \beta}_0(\tau)$.
More conditions are needed.

{\bf Assumption 1} The covariate $X$ satisfies that $\|X_i\|\le M$ for some constant $M$ uniformly in $i=1,\ldots, n$.

{\bf Assumption 2} The function $\dot{{\bm \beta}}(\tau)$ is bounded away from $0$ for all $\epsilon \le \tau \le 1-\epsilon$ and $0<\epsilon<1$.

{\bf Assumption 3} The bandwidth $h$ for the derivative estimation satisfies $h=o(n^{-\delta})$ with $0<\delta<1/2$.

\vspace{0.2in}
Hereafter, mathematic operators of vectors (matrices) $A$ and $B$, such as $A+B$ and $A/B$,  stand for the corresponding operators of each component of $A$ and $B$.

\noindent  {\bf Proof of Theorem 1. }  We prove Theorem 1 in the several steps.

\noindent {\it Step 1. } To prove
\begin{align}
\label{den}
\sup_{-M_{\epsilon}<t<M_{\epsilon}} \left| \hat f_{Y|X}(t) -f_{Y|X}(t)  \right| = O_p\left(\frac{1}{\sqrt{nh^2}}+ \frac{\{log(n)\}^{3/2}}{nh}+h^2\right),
\end{align}
where $\hat f_{Y|X}(t) = 1/ \{X^\top \hat{\dot{{\bm \beta}}}(\tau)\}$ with $t = X^\top {\bm \beta}(\tau)$ for any fixed $\epsilon\le \tau\le 1-\epsilon$,
\begin{align}
\label{estden}
\hat{\dot{{\bm \beta}}}(\tau) \equiv \frac{\hat {{\bm \beta}}^c(\tau+h) - \hat {{\bm \beta}}^c(\tau-h) }{2h},
\end{align}
and $M_\epsilon$ is certain constant large enough depending on $\epsilon$ and $M$.
To this end, first, standard approximation using Taylor expansion shows that
\begin{align}
\label{app1}
{\bm \beta}_0(\tau+h)-{\bm \beta}_0(\tau-h) &= [{\bm \beta}_0(\tau+h)-{\bm \beta}_0(\tau)] -  [{\bm \beta}_0(\tau-h)-{\bm \beta}_0(\tau)] \nonumber \\
&= \dot{{\bm \beta}}_0(\tau)\times 2h + O(h^3),
\end{align}
which implies
\begin{align}
\label{app2}
\frac{{\bm \beta}_0(\tau+h)-{\bm \beta}_0(\tau-h)}{2h}&= \dot{{\bm \beta}}_0(\tau) + O(h^2).
\end{align}
Next, by a result in Portnoy(2012, page 1733),    we have
\begin{align}
\label{portnoy}
\hat {{\bm \beta}}^c(\tau) - {\bm \beta}_0(\tau) &= O_p\left(\frac{1}{\sqrt{n}}+\frac{(\log{n})^{3/2}}{n} \right), \nonumber \\
\hat{ \dot{{\bm \beta}}}(\tau) - \dot{{\bm \beta}_0}(\tau) &= O_p \left(\frac{1}{\sqrt{n h^2}}+\frac{(\log{n})^{3/2}}{n h} + h^2 \right),
\end{align}
uniformly for all $\epsilon \le \tau \le 1-\epsilon$.
A straightforward calculation yields that
\begin{align}
\label{inv2}
\frac{1}{\hat { \dot{{\bm \beta}}}(\tau)} =  \frac{1}{ { \dot{{\bm \beta}_0}}(\tau) } - \frac{\hat{\dot{{\bm \beta}}}(\tau)- { \dot{{\bm \beta}_0}}(\tau) }{ \left[ { \dot{{\bm \beta}_0}}(\tau) \right]^2} + \frac{\left[\hat{\dot{{\bm \beta}}}(\tau)- { \dot{{\bm \beta}_0}}(\tau) \right]^2}{ \left[ { \dot{{\bm \beta}}}(\tau) \right]^2 \times \hat { \dot{{\bm \beta}}}(\tau)}.
\end{align}
Under condition (A2),  $\inf_{\epsilon <\tau< 1-\epsilon} \hat {\dot {\bm \beta}} (\tau)>c>0 $ for some positive constant $c$.
Together with 
(\ref{portnoy}) and
(\ref{inv2}), we have
 \begin{align}
\label{dbeta2}
\sup_{\epsilon<\tau <1-\epsilon} \left | \frac{1}{\hat{ \dot{{\bm \beta}}}(\tau)} - \frac{1}{\dot{{\bm \beta}_0}(\tau)} \right | &= O_p \left(\frac{1}{\sqrt{n h^2}}+\frac{(\log{n})^{3/2}}{n h} + h^2 \right).
\end{align}
Next, by the boundedness of $X$ in assumption (A1), we have
 \begin{align}
\label{dbeta3}
\sup_{\epsilon<\tau <1-\epsilon} \left | \frac{1}{X^\top \hat{ \dot{{\bm \beta}}}(\tau)} - \frac{1}{X^\top \dot{{\bm \beta}_0}(\tau)} \right | &= O_p \left(\frac{1}{\sqrt{n h^2}}+\frac{(\log{n})^{3/2}}{n h} + h^2 \right),
\end{align}
which implies
\begin{align*}
\label{den2}
\sup_{-M_{\epsilon}<t<M_{\epsilon}} \left| \hat f_{Y|X}(t) -f_{Y|X}(t)  \right| = O_p\left(\frac{1}{\sqrt{nh^2}}+ \frac{\{log(n)\}^{3/2}}{nh}+h^2\right).
\end{align*}

\vspace{0.1in}

\noindent {\it Step 2. } To prove
\begin{align*}
\sup_{1 \le l\le L} \left| \hat {\bm D}_l- {\bm D}_l \right|= O_p \left(\frac{1}{\sqrt{nh^2}}+ \frac{\{log(n)\}^{3/2}}{nh}+h^2 \right ).
\end{align*}
To this end, first, we need to evaluate the order of  $ | {1}/{\{X^\top \hat{ \dot{{\bm \beta}}}(\tau)\}^2} - {1}/{\{X^\top \dot{{\bm \beta}_0}(\tau)\}^2} |$ uniformly in $\tau \in (\epsilon, 1-\epsilon)$. We show that
\begin{align}
\frac{1}{\left\{\hat { \dot{{\bm \beta}}}(\tau)\right\}^2} - \frac{1}{ \left\{{ \dot{{\bm \beta}_0}}(\tau)\right\}^2 } =
\frac{\{{ \dot{{\bm \beta}_0}}(\tau)+\hat{\dot{{\bm \beta}}}(\tau)\}\{{\dot{{\bm \beta}_0}}(\tau)-\hat{\dot{{\bm \beta}}}(\tau)\}}{ \left\{ { \dot{{\bm \beta}_0}}(\tau) \right\}^2
\left\{\hat { \dot{{\bm \beta}}}(\tau)\right\}^2}.
\end{align}
Under Assumption 2, $\inf_{\epsilon <\tau< 1-\epsilon}  {\dot {\bm \beta}} (\tau)>0 $ and $\inf_{\epsilon <\tau< 1-\epsilon} \hat {\dot {\bm \beta}} (\tau)>c>0 $ for some positive constant $c$. Then,
 \begin{align}
 \label{theta22}
\sup_{\epsilon<\tau <1-\epsilon}\left | \frac{1}{\left\{\hat { \dot{{\bm \beta}}}(\tau)\right\}^2} - \frac{1}{ \left\{{ \dot{{\bm \beta}_0}}(\tau)\right\}^2 }\right |
&= C\sup_{\epsilon<\tau <1-\epsilon} \| \hat{\dot{{\bm \beta}}}(\tau) - \dot{{\bm \beta}_0}(\tau)  \| \nonumber \\
&= O_p\left(\frac{1}{\sqrt{nh^2}}+ \frac{\{log(n)\}^{3/2}}{nh}+h^2\right),
\end{align}
for $n$ large enough. For brevity,
we denote $\theta_l\equiv  \dot{{\bm \beta}_0}(\tau_l) $ and $\hat\theta_l\equiv  \hat{\dot{{\bm \beta}}}(\tau_l)$.  Write, for any $l=2,\ldots,L+1$,
\begin{align*}
\frac{1}{\hat \theta_l \hat \theta_{l-1}} - \frac{1}{\theta_l \theta_{l-1}}= \frac{\theta_{l-1}- \hat \theta_{l-1} }{\hat \theta_l \hat\theta_{l-1} \theta_{l-1}}+ \frac{\theta_l - \hat\theta_l}{\hat\theta_l \theta_l \theta_{l-1}} .
\end{align*}
Under Assumption (A2), one can check that
\begin{align*}
\sup_{2 \le l\le (L+1)} \left | \frac{1}{\hat \theta_l \hat \theta_{l-1}} - \frac{1}{\theta_l \theta_{l-1}}  \right |&= C\{
\sup_{2 \le l\le (L+1)}\|\theta_{l-1}-\hat\theta_{l-1} \|+ \sup_{1 \le l\le L }\|\theta_l- \hat\theta_l \|\}
\end{align*}
for $n$ large enough.
Hence,
\begin{align*}
\sup_{ 2 \le l\le (L+1)}\left| \frac{1}{\hat{\dot{\bm \beta}}(\tau_l) \hat{\dot{\bm \beta}}(\tau_{l-1})} - \frac{1}{{\dot{\bm \beta}_0}(\tau_l) {\dot{\bm \beta}_0}(\tau_{l-1})}  \right| = O_p (\frac{1}{\sqrt{nh^2}}+ \frac{\{log(n)\}^{3/2}}{nh}+h^2 ).
\end{align*}
Given that $X$ is bounded, we have
\begin{align}
\label{theta4}
&\sup_{ 2 \le l\le (L+1) }\left| \frac{1}{X^\top \hat{\dot{\bm \beta}}(\tau_l)\times X^\top \hat{\dot{\bm \beta}}(\tau_{l-1})} - \frac{1}{{X^\top \dot{\bm \beta}_0}(\tau_l)\times {X^\top \dot{\bm \beta}_0}(\tau_{l-1})}  \right|\nonumber\\
 =& O_p \left(\frac{1}{\sqrt{nh^2}}+ \frac{\{log(n)\}^{3/2}}{nh}+h^2 \right ).
\end{align}
From the definition matrices ${\bm A}$ and ${\bm B}$ in section 2, combining  (\ref{theta22}) and (\ref{theta4}),  we have
\begin{align}
\label{dsup}
\sup_{1 \le l\le L} \left| \hat {\bm D}_l- {\bm D}_l \right|= O_p \left(\frac{1}{\sqrt{nh^2}}+ \frac{\{log(n)\}^{3/2}}{nh}+h^2 \right ).
\end{align}

\vspace{0.1in}

{\it Step 3. } For ease of presentation, let $\hat \eta_l\equiv 1/X^\top \hat \theta_l$
and  $\eta_l\equiv 1/X^\top \theta_l$. Recall that $\theta_l\equiv  \dot{{\bm \beta}_0}(\tau_l) $ and $\hat\theta_l\equiv  \hat{\dot{{\bm \beta}_0}}(\tau_l)$.
Consider
\begin{align*}
&\left| \eta_{l}{\bm D}_{l}-\hat\eta_{l}\hat {\bm D}_{l} \right| \nonumber \\
 &= \left| \eta_{l}{\bm D}_{l}- (\hat\eta_{l}- \eta_{l}+ \eta_{l})(\hat {\bm D}_{l}-
{\bm D}_{l} + {\bm D}_{l})  \right| \nonumber \\
&= \left| \eta_{l}{\bm D}_{l}- \{(\hat\eta_{l}-\eta_{l})(\hat {\bm D}_{l}-
{\bm D}_{l})+\eta_{l}(\hat {\bm D}_{l}-
{\bm D}_{l})+ {\bm D}_{l}( \hat\eta_{l}-\eta_{l}) +\eta_{l}{\bm D}_l\} \right| \nonumber \\
&= \left|-(\hat\eta_{l}-\eta_{l})(\hat {\bm D}_{l}-
{\bm D}_{l})-\eta_l(\hat {\bm D}_{l}-
{\bm D}_{l}) -{\bm D}_l(\hat\eta_{l}-\eta_{l})   \right|.
\end{align*}
Hence,
\begin{align*}
&\sup_{2\le l\le L+1} \left| \eta_l {\bm D}_l-\hat \eta_{l-1} \hat {\bm D}_{l-1} \right| \nonumber \\
&= \sup_{2\le l\le L+1} \left| (\hat\eta_{l}-\eta_{l})(\hat {\bm D}_{l}-
{\bm D}_{l})+\eta_l(\hat {\bm D}_{l}-
{\bm D}_{l}) +{\bm D}_l(\hat\eta_{l}-\eta_{l})  \right|
\end{align*}
Using (\ref{dbeta2}) and (\ref{dsup}), we have shown
\begin{align*}
&\sup_{2\le l\le L+1} \left| \eta_l {\bm D}_l-\hat \eta_{l-1} \hat {\bm D}_{l-1} \right|  = O_p \left(\frac{1}{\sqrt{nh^2}}+ \frac{\{log(n)\}^{3/2}}{nh}+h^2 \right ).
\end{align*}

{\it Step 4.} To evaluate the estimated score function $\hat S_k(y,x)$ in Proposition 1 by plugging in $\hat f_{Y|X}(\cdot)$, $\hat {\bm D}_l$ and $\hat {\bm \beta}^c(\tau_l)$, $l=1,\ldots, L$ into $S_k(y,x)$. To be concise, we define
\begin{align}
\hat {S}_k(y_i,x_i)\equiv \sum_{l=1}^{L+1} \frac{\hat a_{l-1}-\hat a_l}{\tau_l-\tau_{l-1}} \left[ I\{x_i^\top \hat{{\bm \beta}}^c(\tau_{l-1})< y_i< x_i^\top \hat{{\bm \beta}}^c(\tau_l) \} -(\tau_l-\tau_{l-1}) \right],
\end{align}
where $\hat a_l=\hat \eta_l \hat {\bm D}_l$ and $a_l=\eta_l {\bm D}_l$,  $\hat\eta_l$. 
First, write
\begin{align}
\label{al}
\frac{\hat a_{l-1}-\hat a_l}{\tau_l-\tau_{l-1}}= \frac{ a_{l-1}- a_{l}}{\tau_l-\tau_{l-1}} +\frac{\hat a_{l-1}- a_{l-1}- (\hat a_l- a_l)}{\tau_l-\tau_{l-1}}.
\end{align}
We also write
\begin{align}
\label{ihat}
 I\{x_i^\top \hat{{\bm \beta}}^c(\tau_{l-1}) < y_i< x_i^\top \hat{{\bm \beta}}^c(\tau_l) \} =  I\{y_i> x_i^\top \hat {\bm \beta}^c(\tau_{l-1})\}-I\{y_i> x_i^\top \hat {\bm \beta}^c(\tau_l)\}.
\end{align}
We first consider the first term in (\ref{ihat}) and write
\begin{align}
 I\{y_i> x_i^\top \hat {\bm \beta}^c(\tau_{l-1})\}&=  I\{y_i> x_i^\top \hat {\bm \beta}^c(\tau_{l-1})\} -  I\{y_i> x_i^\top {\bm \beta}_0(\tau_{l-1})\} + I\{y_i> x_i^\top {\bm \beta}_0(\tau_{l-1})\}  \nonumber \\
 &= I\{x_i^\top {\bm \beta}_0(\tau_{l-1})< y_i< x_i^\top \hat {\bm \beta}^c(\tau_{l-1})  \} +  I\{y_i> x_i^\top {\bm \beta}_0(\tau_{l-1})\} \nonumber  \\
 &\equiv \hat\Delta_{l-1}^i + I\{y_i> x_i^\top {\bm \beta}_0(\tau_{l-1})\}
 \end{align}
Based on the fact that $Var(\hat\Delta^i_{l-1})= O( \frac{1}{\sqrt{n h^2}}+\frac{(\log{n})^{3/2}}{n h} + h^2 )$, we can show
\begin{align}
\sup_{1\le l\le L} |\hat \Delta_l^i | = O_p\left(\sqrt{\frac{1}{\sqrt{n h^2}}+\frac{(\log{n})^{3/2}}{n h} + h^2 } \right).
\end{align}
Then, by the monotonicity implied by the quantile regression model,
\begin{align}
\label{ihat2}
I\{x_i^\top \hat{{\bm \beta}}^c(\tau_{l-1}) < y_i< x_i^\top \hat{{\bm \beta}}^c(\tau_l) \}= \hat\Delta_{l-1}^i-\hat \Delta_l^i+ I\{x_i^\top {{\bm \beta}_0}(\tau_{l-1}) < y_i<x_i^\top {{\bm \beta}_0}(\tau_l) \}
\end{align}
Using (\ref{al}) and (\ref{ihat2}), we have
\begin{align}
&\ \ \hat {S}_k(y_i,x_i) \nonumber \\
& \equiv \sum_{l=1}^{L+1} \left\{\frac{a_{l-1}-a_l}{\tau_l-\tau_{l-1}}+\frac{(\hat a_{l-1}- a_{l-1})-(\hat a_l-a_l)}{ \tau_l-\tau_{l-1}}  \right\} \times \nonumber \\
& \hskip 1cm \left[ \hat\Delta^i_{l-1}-\hat \Delta^i_l + I\{x_i^\top {{\bm \beta}_0}(\tau_{l-1})< y_i < x_i^\top {{\bm \beta}_0}(\tau_l) \} -(\tau_l-\tau_{l-1}) \right], \nonumber \\
&= \sum_{l=1}^{L+1} \left\{\frac{a_{l-1}-a_l}{\tau_l-\tau_{l-1}}\right\}\left[\hat \Delta^i_{l-1}-\hat\Delta^i_l + I\{x_i^\top {{\bm \beta}_0}(\tau_{l-1})< y_i < x_i^\top {{\bm \beta}_0}(\tau_l) \} -(\tau_l-\tau_{l-1}) \right], \nonumber  \\
&+ \sum_{l=1}^{L+1} \left\{ \frac{(\hat a_{l-1}- a_{l-1})-(\hat a_l-a_l)}{ \tau_l-\tau_{l-1}}\right\} \left[ \hat\Delta^i_{l-1}-\hat\Delta^i_l + I\{x_i^\top {{\bm \beta}_0}(\tau_{l-1})< y_i < x_i^\top {{\bm \beta}_0}(\tau_l) \} -(\tau_l-\tau_{l-1}) \right], \nonumber  \\
&= \sum_{l=1}^{L+1} \frac{a_{l-1}-a_l}{\tau_l-\tau_{l-1}}\times \left[I\{x_i^\top {{\bm \beta}_0}(\tau_{l-1})< y_i < x_i^\top {{\bm \beta}_0}(\tau_l) \} -(\tau_l-\tau_{l-1}) \right]   \nonumber \\
&  +\sum_{l=1}^{L+1}  \frac{(a_{l-1}-a_l)(\hat\Delta^i_{l-1}-\hat\Delta^i_l)}{\tau_l-\tau_{l-1}}+\sum_{l=1}^{L+1} \frac{(\hat a_{l-1}- a_{l-1})-(\hat a_l-a_l)}{ \tau_l-\tau_{l-1}} \times (\hat\Delta^i_{l-1}-\hat\Delta^i_l) \nonumber \\
& + \sum_{l=1}^{L+1} \frac{(\hat a_{l-1}- a_{l-1})-(\hat a_l-a_l)}{\tau_l-\tau_{l-1}}\times \left[I\{x_i^\top {{\bm \beta}_0}(\tau_{l-1})< y_i < x_i^\top {{\bm \beta}_0}(\tau_l) \} -(\tau_l-\tau_{l-1}) \right] \nonumber \\
&\equiv  S_k^0(y_i, x_i)+ \Pi_1+\Pi_2+\Pi_3.
\end{align}
We first consider $\Pi_1$. For $L$ large but fixed, as obviously $E(\hat\Delta^i_l)\neq 0$, then it can be shown that,
\begin{align}
\label{pi1}
\Pi_1 = \sum_{l=1}^{L+1}  \frac{(a_{l-1}-a_l)(\hat\Delta^i_{l-1}-\hat\Delta^i_l)}{\tau_l-\tau_{l-1}} =  O_p\left(L\sqrt{\frac{1}{\sqrt{nh^2}}+ \frac{\{log(n)\}^{3/2}}{nh}+h^2  }\right),
\end{align}
uniformly in $i$, where $1\le k\le L$. Next,
\begin{align}
\label{pi2}
\Pi_2 &= \sum_{l=1}^{L+1} \frac{(\hat a_{l-1}- a_{l-1})-(\hat a_l-a_l)}{\tau_l-\tau_{l-1}} \times (\hat\Delta^i_{l-1}-\hat\Delta^i_l) \nonumber \\
&=  O_p\left(L\left\{\frac{1}{\sqrt{nh^2}}+ \frac{\{log(n)\}^{3/2}}{nh}+h^2 \right\}^\frac{3}{2} \right)
\end{align}
uniformly in $i$. Similarly, it can be shown that
\begin{align}
\label{pi3}
\Pi_3 = O_p\left(L\left\{\frac{1}{\sqrt{nh^2}}+ \frac{\{log(n)\}^{3/2}}{nh}+h^2 \right\} \right)
\end{align}
uniformly in $i$. Consequently, for any $1\le k\le L$,
\begin{align*}
&\left| \frac{1}{n}\sum_{i=1}^n \hat S_k(y_i, x_i)- \frac{1}{n}\sum_{i=1}^n  S_k^0(y_i, x_i) \right|  \nonumber \\
=& \frac{1}{n} \sum_{i=1}^n \left\{\sum_{l=1}^{L+1} \frac{(a_{l-1}-a_l)(\hat\Delta^i_{l-1}-\hat\Delta^i_l)}{\tau_l-\tau_{l-1}}  \right\}
+  \frac{1}{n} \sum_{i=1}^n \left\{\sum_{l=1}^{L+1} \frac{(\hat a_{l-1}- a_{l-1})-(\hat a_l-a_l)}{ \tau_l-\tau_{l-1}} \times (\hat\Delta^i_{l-1}-\hat\Delta^i_l)  \right\}  \nonumber \\
& + \frac{1}{n}\sum_{i=1}^n  \left( \sum_{l=1}^{L+1} \frac{(\hat a_{l-1}- a_{l-1})-(\hat a_l-a_l)}{\tau_l-\tau_{l-1}}\times \left[I\{x_i^\top {{\bm \beta}_0}(\tau_{l-1})< y_i < x_i^\top {{\bm \beta}_0}(\tau_l) \} -(\tau_l-\tau_{l-1}) \right]   \right).
\end{align*}
Combining (\ref{pi1}), (\ref{pi2})  and (\ref{pi3}), we have
\begin{align}
\label{estrue}
\sup_{1\le k\le L}\left| \frac{1}{n}\sum_{i=1}^n \hat S_k(y_i, x_i)- \frac{1}{n}\sum_{i=1}^n  S_k^0(y_i, x_i) \right| = O_p\left(\frac{1}{\sqrt{n}}\left\{\frac{1}{\sqrt{nh^2}}+ \frac{\{log(n)\}^{3/2}}{nh}+h^2 \right\}^{1/2}  \right)
\end{align}
for fixed $L$.

For the variance estimation, since $\sigma^2_{kj}= 1/\{{\bm U}^\top_{kj} {\bm U} {\bm U}_{kj}\}$, we estimate it by the plug-in method.
In view of the fact that
\begin{align*}
\sup_{1 \le l\le L} \left| \hat {\bm D}_l- {\bm D}_l \right|= O_p \left(\frac{1}{\sqrt{nh^2}}+ \frac{\{log(n)\}^{3/2}}{nh}+h^2 \right ),
\end{align*}
we have
\begin{align*}
\sup_{1 \le k\le L,  1\le j\le p} \left| \hat \sigma^2_{kj}-\sigma^2_{kj} \right|= O_p \left(\frac{1}{\sqrt{nh^2}}+ \frac{\{log(n)\}^{3/2}}{nh}+h^2 \right ).
\end{align*}

{\it Step 5. } Lastly, recall the proposed one-step efficient estimation in section 2, that is, for  $j=1,2,\ldots,p$
\begin{align*}
\hat{{\bm \beta}}_{j}(\tau_k)=\hat{ {\bm \beta}}^c_{j}(\tau_k)+\hat \sigma^2_{kj}\frac{\sum_{i=1}^n \hat S_{kj}(y_i,x_i)}{n}.
\end{align*}
For ease of presentation, we define
\begin{align*}
 {S}^0_{k}(y_i,x_i; {\bm \beta}(\cdot))\equiv \sum_{l=1}^{L+1} \frac{a_{l-1}- a_l}{\tau_l-\tau_{l-1}} \left[ I\{x_i^\top {{\bm \beta}}(\tau_{l-1})< y_i< x_i^\top {{\bm \beta}}(\tau_l) \} -(\tau_l-\tau_{l-1}) \right].
\end{align*}
According to the above definition, $S^0_k(y_i,x_i)={S}^0_{k}(y_i,x_i; {\bm \beta}_0(\tau))$.
Observe that, for  $j=1,2,\ldots,p$,
\begin{align*}
\hat \beta_j(\tau_k)&= \beta_{0,j}(\tau_k)+\{\hat{ \beta}^c_{j}(\tau_k) - \beta_{0,j}(\tau_k)  \}+ \{\hat \sigma^2_{kj} - \sigma^2_{kj} +\sigma^2_{kj} \} \nonumber \\
& \hskip 1cm \times \frac{1}{n}\sum_{i=1}^n \{ \hat S_{kj}(y_i,x_i) -  S_{kj} ^0(y_i, x_i) + S_{kj} ^0(y_i, x_i) \},
\end{align*}
which implies
\begin{align*}
&\hat \beta_j(\tau_k)- \beta_{0,j}(\tau_k) \nonumber \\
=& \{\hat{ \beta}^c_{j}(\tau_k) - \beta_{0,j}(\tau_k)  \}+ (\hat \sigma^2_{kj} - \sigma^2_{kj})
\frac{1}{n}\sum_{i=1}^n \{ \hat S_{kj}(y_i,x_i) -  S_{kj} ^0(y_i, x_i)\} \nonumber \\
& +\sigma^2_{kj}  \frac{1}{n}\sum_{i=1}^n \{ \hat S_{kj}(y_i, x_i) -  S_{kj} ^0(y_i, x_i)\}
+ \frac{1}{n}\sum_{i=1}^n S_{kj} ^0(y_i, x_i) (\hat \sigma^2_{kj} - \sigma^2_{kj})\nonumber \\
& +  \sigma^2_{kj} \frac{1}{n}\sum_{i=1}^n S_{kj} ^0(y_i, x_i) .
\end{align*}
 Since $S_{kj}^0$ is the efficient score, one can show that for all $1 \le k\le L$ and $1\le j\le p$,
\begin{align}
\label{effscore}
&\sup_{\|{\bm \beta}(\tau_l)-{\bm \beta}_{0}(\tau_l)\|\le B_n,  1\le l\le L }
\left\{ \frac{\left|\sum_{i=1}^n S_{k}^0(y_i, x_i; {\bm \beta}(\cdot)) - \sum_{i=1}^n S_{k} ^0(y_i, x_i)+n A^k_n \{{\bm \beta}(\tau_k)-{\bm \beta}_0(\tau_k)\} \right| }{\sqrt{n}+ n\|{\bm \beta}(\tau_k)-{\bm \beta}_0(\tau_k)\|^{1+\delta}} \right\} \nonumber \\
=& o_p(1),
\end{align}
where $A^k_n= (1/n)\sum_{i=1}^n\{\partial/(\partial {\bm \beta}(\tau_k)) S_{kj}^0(y_i, x_i; {\bm \beta}(\cdot)) \}$, $0<\delta<1$, and $B_n\to 0$ as $n\to \infty$. One can also see that
\begin{align*}
&\frac{1}{n}\sum_{i=1}^n S_{kj}^0(y_i, x_i; \hat {\bm \beta}^c(\cdot)) - \sum_{i=1}^n S_{kj} ^0(y_i, x_i) \nonumber \\
&=\frac{1}{n}\sum_{i=1}^n \sum_{l=1}^{L+1}b_{l,l-1}(\hat\Delta^i_{l-1}-\hat\Delta^i_l),
\end{align*}
where $b_{l,l-1}\equiv (\hat a_{l-1}-\hat a_l)/(\tau_l-\tau_{l-1})$ and $\hat\Delta^i_l$ is defined earlier. Since
$$Var\left\{ \frac{1}{n}\sum_{i=1}^n \sum_{l=1}^{L+1}b_{l,l-1}(\hat\Delta^i_{l-1}-\hat\Delta^i_l)\right\}= o\left[\frac{n}{n^2} \left\{ \frac{1}{\sqrt{n h^2}}+\frac{(\log{n})^{3/2}}{n h} + h^2\right\} \right], $$
one can show
\begin{align*}
\frac{1}{n}\sum_{i=1}^n \sum_{l=1}^{L+1}b_{l,l-1}(\hat\Delta^i_{l-1}-\hat\Delta^i_l)= o_p(\frac{1}{\sqrt{n}})
\end{align*}
under Assumption (A3).

Second, by the central limit theorem,
\begin{align}
\label{clt}
\frac{1}{\sqrt{n}} \sum_{i=1}^n S_{kj}^0(y_i,x_i) \to N(0, (\sigma^2_{kj})^{-1}),
\end{align}
in distribution as $n\to \infty$.

Third, by the law of large numbers, $-A^k_n\to (\sigma^2_{kj})^{-1}$ in probability as $n\to \infty$. Recall that $\hat \sigma^2_{kj}-\sigma^2_{kj}=o_p(1)$.
Moreover, using (\ref{estrue}) and (\ref{effscore}), we have
\begin{align*}
\hat {\bm \beta}^c(\tau_k) -{\bm \beta}_0(\tau_k)+ \sigma^2_{kj}A^k_n\{ \hat {\bm \beta}^c(\tau_k) -{\bm \beta}_0(\tau_k)\}=o_p(n^{-\frac{1}{2}}).
\end{align*}

Finally,
\begin{align*}
\hat \beta_j(\tau_k)=\beta_{0,j}(\tau_k)+ \sigma^2_{kj} \frac{1}{n}\sum_{i=1}^n S^0_{kj}(y_i, x_i)+r_n,
\end{align*}
where $r_n= O_p(1/\sqrt{n^{3/2}h}+h/\sqrt{n})$ for $n$ sufficiently large.  In view of (\ref{clt}) and Assumption (A3), we have shown
\begin{align*}
\sqrt{n}\{\hat \beta_j(\tau_k)-\beta_{0,j}(\tau_k) \} \to N(0, \sigma^2_{kj})
\end{align*}
in distribution as $n\to \infty$. We complete the proof of Theorem 1.

\vspace{0.1in}

{\it Remark: } Note that we need  $r_n=o_p({1}/{\sqrt{n}})$ to ensure the asymptotic normality, which requires $$\frac{1}{n^{\frac{1}{4}} h^{\frac{1}{2}}}+h \to 0$$ as $n\to \infty$. That is to say, we need to assume
$nh^2 \to \infty$ and $h\to 0$, that is $h=o(n^{-\delta})$ with $0<\delta<1/2$.

\end{document}